\documentclass[
    aps,
    physrev,
    reprint,
    superscriptaddress,
]{revtex4-2}

\usepackage{graphicx}
\usepackage[normalem]{ulem}
\usepackage{hyperref}

\hypersetup{
    colorlinks = true,
    linkcolor = blue,
    citecolor = blue,
    urlcolor = blue,
}

\newcommand{\Tc}{T$_\textrm{c}$}
\newcommand{\Bc}{B$_\textrm{cool}$}
\newcommand{\FCGT}{(Fe$_{0.5}$Co$_{0.5}$)$_5$GeTe$_2$}
\newcommand{\FGT}{Fe$_5$GeTe$_2$}

\begin{document}

\title{Mapping metastable magnetic textures in \texorpdfstring{\FCGT{}}{(Fe0.5Co0.5)5GeTe2} with in-situ Lorentz transmission electron microscopy}

\author{Reed Yalisove}
\affiliation{University of California, Berkeley, Department of Materials Science and Engineering, Berkeley, CA, United States}
\affiliation{Lawrence Berkeley National Lab, National Center for Electron Microscopy, Berkeley, CA, United States}

\author{Hongrui Zhang}
\affiliation{University of California, Berkeley, Department of Materials Science and Engineering, Berkeley, CA, United States}

\author{Xiang Chen}
\affiliation{University of California, Department of Physics, Berkeley, CA, United States}

\author{Fanhao Meng}
\author{Jie Yao}
\affiliation{University of California, Berkeley, Department of Materials Science and Engineering, Berkeley, CA, United States}
\affiliation{Lawrence Berkeley National Lab, Materials Sciences Division, Berkeley, CA, United States}

\author{Robert Birgeneau}
\affiliation{Lawrence Berkeley National Lab, Materials Sciences Division, Berkeley, CA, United States}
\affiliation{University of California, Department of Physics, Berkeley, CA, United States}

\author{Ramamoorthy Ramesh}
\affiliation{University of California, Berkeley, Department of Materials Science and Engineering, Berkeley, CA, United States}

\author{Mary C. Scott}
\affiliation{University of California, Berkeley, Department of Materials Science and Engineering, Berkeley, CA, United States}
\affiliation{Lawrence Berkeley National Lab, National Center for Electron Microscopy, Berkeley, CA, United States}

\date{\today}

\begin{abstract}
Topologically protected magnetic textures are a promising route to low-energy control of magnetism, but they are most often studied away from ambient conditions, typically at low temperatures and high magnetic fields. Here we use in-situ Lorentz transmission electron microscopy with control of temperature and magnetic field to investigate the skyrmion metastability in \FCGT{} (FCGT). By field-cooling FCGT in magnetic fields of different magnitude to different base temperatures and then removing the applied field, we create meta(stable) zero-field magnetic states. We use this method to build a phase diagram of the zero-field metastable spin structures in FCGT, which will be critical for selecting the desired topologically-protected spin state for future studies to manipulate magnetism with stimuli such as electric current, electric field, mechanical strain, and more.
\end{abstract}

\maketitle

\section*{Introduction}
Magnetic skyrmions and other complex, topologically-protected spin structures are an exciting medium for designing new means to control magnetism---via electric current, electric field, mechanical strain, and more \cite{jonietz_spin_2010, jiang_blowing_2015, jiang_direct_2017, schott_skyrmion_2017, shibata_large_2015, finazzi_laser-induced_2013}. Reliable low-energy control of magnetism could decrease the energy demands of current magnetic-domain based information storage systems \cite{bocdanov_properties_1994, fert_skyrmions_2013}. Within the classical framework \cite{kittel_charles_introduction_2004},  domain structures arise from the competition between the exchange interaction, magnetic anisotropy and the minimization of magnetostatic energy. The quantum mechanical spin-orbit effect is introduced in a second-order expansion of the exchange interaction Hamiltonian (Dzyaloshinskii–Moriya interaction, DMI) and is only accessible in systems with bulk or interfacial inversion-symmetry breaking \cite{dzyaloshinsky_thermodynamic_1958, moriya_anisotropic_1960}. 

\begin{figure}
    \centering
    \includegraphics[width=\linewidth]{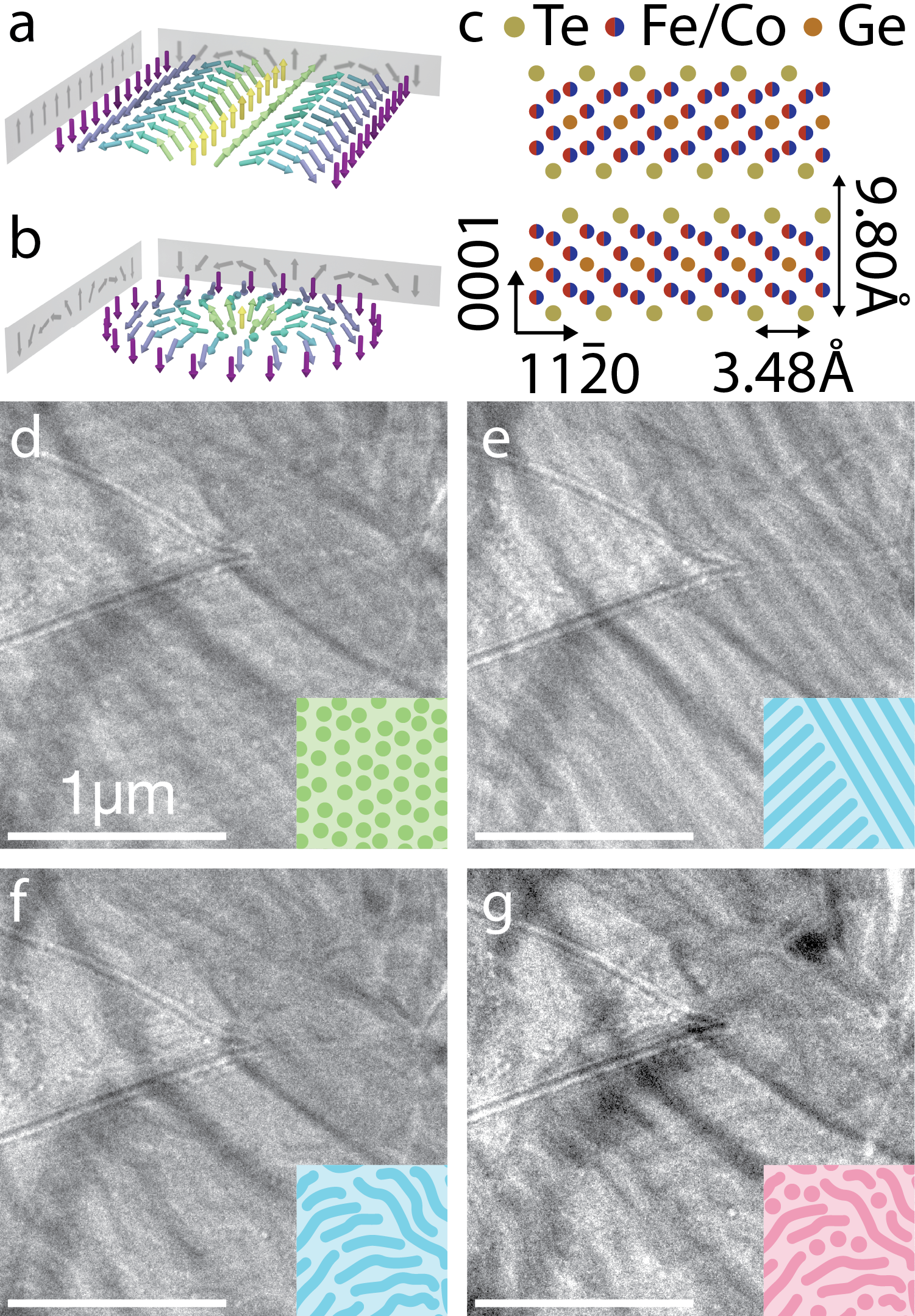}
    \caption{Magnetic textures in FCGT. Spin arrangement of (a) N\'eel skyrmions and (b) cycloidal states. In (a) and (b) the gray panels show the spin alignments along a slice through the center of the spin textures. c) Crystal structure  of Van-der-Waals layered FCGT. d-g) LTEM images of FCGT showing (d) skyrmions, (e) parallel cycloidal domains, (f) labyrinthine cycloidal domains, and (g) mixed skyrmion-cycloidal state with insets illustrating the expected appearance of each state.}
    \label{fig:states}
\end{figure}

The DMI and other non-classical spin-spin interactions such as 3- or 4-spin exchange interactions lead to new spin structures including magnetic helices, cycloids, and skyrmions \cite{nagaosa_topological_2013}. The anti-symmetric DMI causes the spin direction to cant from one atom to the next, creating a helix or cycloid of spin directions across the material as shown in Figure \ref{fig:states}a \cite{dzyaloshinskii_theory_1964, ishikawa_helical_1976, uchida_real-space_2006}. Helical and cycloidal magnetic textures can have a pitch ranging from 2nm-200nm depending on the strength of the DMI \cite{dzyaloshinskii_theory_1964, ding_observation_2022}. At zero applied magnetic field, the cycloidal (or helical) state is the low-energy spin configuration for materials with a sufficiently strong DMI \cite{rosler_chiral_2011}. Going forward, we refer only to spin cycloids and cycloidal spin states, though the same phenomena can lead to the formation of spin helices and helical spin states \cite{kezsmarki_ne-type_2015}.

Under a perpendicular magnetic field, a skyrmion lattice state can become the low-energy spin configuration. In a discrete skyrmion, the spins rotate 180$^\circ$ from the center to the edge along any radial cross-section (Fig. \ref{fig:states}b) \cite{skyrme_unified_1962, bogdanov_thermodynamically_1994, muhlbauer_skyrmion_2009}. Skyrmions have been reported in MnSi, FeCoSi, FeGe, Fe$_\textrm{x}$GeTe$_2$, FeGd, various heterostructures, and related materials at low temperatures under an applied magnetic field \cite{muhlbauer_skyrmion_2009, yu_real-space_2010,yu_near_2011, ding_observation_2020, zhang_room-temperature_2022, montoya_tailoring_2017, wu_ne-type_2020}. Below their Curie temperatures and at or near zero magnetic field, these materials naturally form cycloidal or helical states. In a higher magnetic field (but below the saturation magnetization) they switch to a skyrmion state---this requires the spins to flip, consistent with a first-order phase transition \cite{muhlbauer_skyrmion_2009, rosler_spontaneous_2006}. No smooth transition of spin arrangement can create or destroy a skyrmion, so it is topologically protected \cite{nagaosa_topological_2013}. For most materials, removing the magnetic field will return the system to the cycloidal state---there is no thermodynamic energy barrier to flipping spins \cite{nagaosa_topological_2013}. 

In some magnetic materials and at some temperatures, non-equilibrium local energy minima exist. A system trapped in one of these local energy minima is metastable \cite{arrhenius_uber_1889}. The equilibrium magnetic state depends on the applied magnetic field, and the energy barrier to flipping spins (and escaping the local minimum) depends on temperature. Practically we can force a system into a metastable magnetic state by choosing a state at high temperature where the barrier to switching is low, then lowering the temperature of the system until it cannot overcome the energy barrier to switch states. 

The metastable zero-field skyrmion state is accessible by cooling a material from its Curie temperature in a magnetic field. Once the cooling field is removed, skyrmions can persist in a metastable state. Metastable skyrmions are a product of the thermal and magnetic field history of the material. Metastable zero-field (and sometimes room temperature) skyrmions have been reported, but the thermal and magnetic field history that yields these metastable states is often not reported or studied \cite{karube_robust_2016, gallagher_robust_2017, zhang_room-temperature_2022}. Here we systematically study the effects of thermal and magnetic field history on complex, zero-field spin structures in order to learn what processing is required to generate metastable spin structures and the conditions in which those spin structures will remain metastable. The thermal and magnetic field history of a material can be described by its path through a temperature-magnetic field phase space. Any difference in final spin state from different paths with the same end point (path dependence) will indicate the presence of metastable states. 

Co-doped \FGT{} is a Van-der-Waals layered material with an antiferromagetic-to-ferromagnetic transition at 50\% Co-doping \cite{zhang_room_2022}. Each layer consists of diagonal rows of Fe or Co sites surrounding a layer of Ge atoms and sandwiched between rows of Te atoms. The bottom row of Te atoms are aligned between layers, giving AA' stacking (Fig. \ref{fig:states}c). Ferromagnetic noncentrosymmetric \FCGT{} (FCGT) hosts metastable skyrmions at room-temperature and zero magnetic field, making it a model system for studying the metastability of magnetic textures \cite{zhang_room-temperature_2022, meisenheimer_ordering_2023}.

We use Lorentz transmission electron microscopy (LTEM) to image the magnetic textures \cite{hale_magnetic_1959, boersch_elektronenmikroskopische_1959}. Figure \ref{fig:states}d-g shows the appearance of different FCGT spin states in LTEM, including the skyrmion state, parallel cycloidal state, labyrinthine cycloidal state, and a mixed skyrmion-cycloidal state. The parallel cycloidal state (Fig. \ref{fig:states}e) has highly linear cycloidal stripes which lie parallel to one another. In the labyrinthine cycloidal state the cycloidal stripes maintain a similar period but bend and meander out of linear, parallel alignment (Fig. \ref{fig:states}f). This distinction has been explored in classical ferrimagnets and chiral helimagnets \cite{molho_irreversible_1987, karube_disordered_2018}. Neutron diffraction and transport measurements are routinely used to identify the skyrmion state (Fig. \ref{fig:states}d) or the cycloidal state (Fig. \ref{fig:states}e) \cite{muhlbauer_skyrmion_2009, jonietz_spin_2010}. However, to distinguish between the parallel cycloidal state and the labyrinthine cycloidal state (Fig. \ref{fig:states}f) with neutron diffraction requires extensive verification \cite{karube_disordered_2018}, and may not be possible with transport measurements. Further, the skyrmions in the aperiodic mixed state (Fig. \ref{fig:states}g) will be invisible to neutron diffraction and difficult to detect with transport measurements. Magnetic force microscopy (MFM) is also a powerful tool for real-space imaging of spin textures, but temperature- and magnetic field-dependent experiments are challenging due to the magnetized scanning tip and slow scan rate. LTEM is thus an ideal platform for real-space imaging of spin states with precise control of temperature and applied magnetic field. 

\section*{Results}
Domain walls in FCGT are N\'eel type, which can be viewed in LTEM by defocusing the electron beam and tilting the sample (we use +600$\mu$m defocus and +18$^\circ$ $\alpha$-tilt throughout this study) \cite{benitez_magnetic_2015}. In order to prevent magnetic saturation of FCGT samples resting in the $\sim$2T magnetic field of the TEM objective lens (FCGT has a coercive field around 0.1T at room temperature), the objective lens is turned off and the electron beam is focused with adjacent condenser lenses that do not impart a magnetic field on the sample. Fine control of small currents in the objective lens is used to probe the sample with a predictable out-of-plane magnetic field, calibrated with a hall sensor attached to a side-entry TEM holder. 

Temperature control is achieved by using a variety of \textit{in-situ} TEM holders. To image FCGT above room temperature, we use a Gatan 652 heating holder, which uses resistive elements to heat a sample. For low-temperature imaging we use a Gatan 636 cooling holder, which has a small liquid-nitrogen-filled dewar that sits outside of the microscope and is thermally coupled to the sample. A resistive heater near the sample in the cooling holder allows for intermediate-temperature data to be collected. Both of these holders can heat FCGT above its Curie temperature ($\sim$390K), which is critical for collecting field-cooling data.

\begin{figure}
    \includegraphics[width = \linewidth]{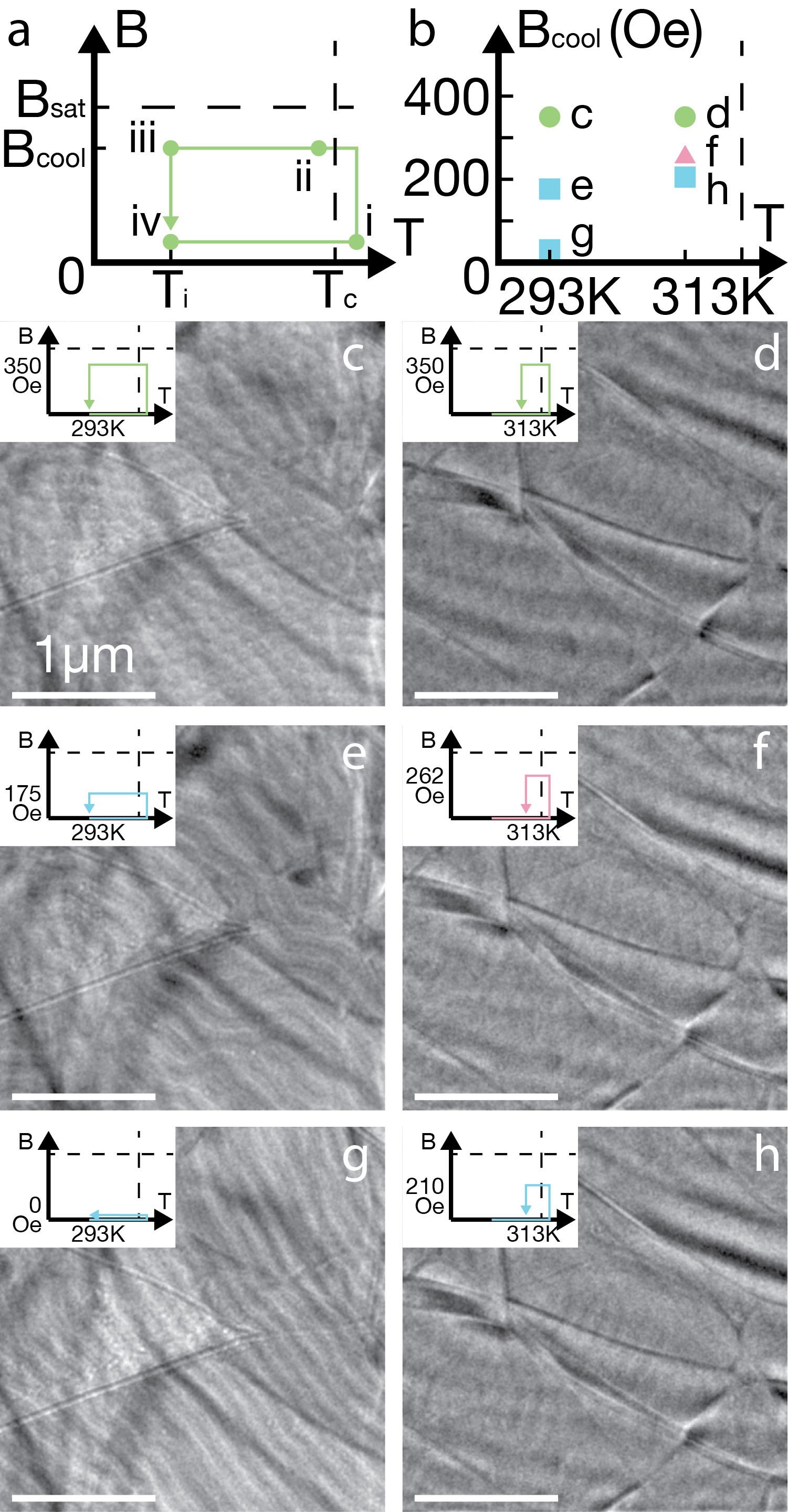}
    \caption{Field-cooling cycles in FCGT. a) Path through temperature-magnetic field phase space for a single field-cooling cycle. On this path we heat the sample above its Curie temperature (i), place it in a magnetic field (ii), cool it to a temperature of interest T$_i$ (iii) and remove the applied magnetic field (iv). (b) A representation of several field-cooling paths (corresponding to [c-h]) with final temperature plotted on the x-axis and cooling field plotted on the y-axis. The shape and color of the plot points corresponds to the resulting magnetic state. c-h) LTEM images of the magnetic textures resulting from a variety of field-cooling paths (inset in each image).}
    \label{fig:loops}
\end{figure}

The various spin textures in Figure \ref{fig:states} exist in FCGT at zero magnetic field after field-cooling along different paths through temperature-magnetic field phase space, indicating several metastable local minima in the energy landscape. When a sample is heated above T$_\textrm{c}$ (point i in Fig. \ref{fig:loops}a), magnetic ordering disappears and the system is in a uniform paramagnetic state. When the system is cooled slightly below T$_\textrm{c}$ (point ii in Fig. \ref{fig:loops}a) magnetic ordering returns and the system enters one of several states depending on the applied magnetic field (B$_\textrm{cool}$). This state tends to persist as the sample is cooled to a temperature (T$_i$) of interest (ii$\rightarrow$iii in Fig. \ref{fig:loops}a) and the magnetic field is removed (iii$\rightarrow$iv in Fig. \ref{fig:loops}a), creating a metastable zero-field spin texture. We represent various paths through the temperature-magnetic field phase space by plotting them with a base temperature, T$_i$, on the x-axis and magnetic field, B$_\textrm{cool}$, on the y-axis, as in Figure \ref{fig:loops}b. The metastable zero-field spin texture imaged with LTEM in Figure \ref{fig:loops}c was created by following the path inset in Figure \ref{fig:loops}c, corresponding to point c in Figure \ref{fig:loops}b. The paths that select the magnetic textures in Figure \ref{fig:loops}d-h are similarly represented by points d-h in Figure \ref{fig:loops}b and we inset the field cooling path on each LTEM image. This scheme lets us create a roadmap through temperature-magnetic field phase space to reach a desired (meta)stable zero-field spin state at any temperature. We measured field cooling loops at temperatures ranging from 203K to 323K. 

\subsection*{Selecting spin texture with field cooling}
We can select a metastable spin state by controlling the field-cooling field (\Bc) and the temperature. The FCGT spin state is reset by heating above \Tc{} to the paramagnetic state. Cooling from \Tc{} to room temperature in zero magnetic field leads to a highly parallel cycloidal state (Fig. \ref{fig:loops}g). This cycloidal state is the minimum energy state for FCGT at room temperature and zero field \cite{zhang_room_2022}. Field cooling with small \Bc{} also yields a cycloidal state (Fig. \ref{fig:loops}e), but the ferromagnetically aligned domains are shorter and less parallel than in the zero-field-cooled cycloidal state. This labyrinthine cycloidal state does not relax to the parallel cycloidal state without heating above the Curie temperature. Molho \textit{et al.} has reported a similar path-dependence on the creation of parallel and labyrinthine magnetic states \cite{molho_irreversible_1987}. Further, the labyrinthine cycloidal state can also be created by applying a magnetic field higher than the magnetic saturation field and then returning to zero field, as discussed later. Metastable room temperature, zero-field magnetic skyrmions are created by cooling with a \Bc{} above a critical field of approximately 210 Oe (Fig \ref{fig:loops}c). These skyrmions are tightly packed and extend across the sample, but they do not appear to have long-range hexagonal order as shown in the Supplemental Material \cite{supp} (Fig. S1). Zero-field mixed states (Fig. \ref{fig:states}g) can be selected by choosing an intermediate (about 210-265 Oe) cooling field. Skyrmions and labyrinthine cycloidal domains coexist at zero magnetic field in the mixed state. Field cooling loops with slightly higher cooling magnetic fields yield the fully skyrmionic spin state, and slightly lower fields yield the fully cycloidal spin state. 

This set of spin states is available from 253K up to 313K with similar cooling magnetic fields. Zero-field cooling selects a parallel cycloidal spin state, field cooling in a small (up to 210 Oe) field selects the labyrinthine cycloidal spin state (Fig. \ref{fig:loops}h), and field cooling in fields between 210 Oe and the saturation field selects the skyrmion spin state (Fig. \ref{fig:loops}d). We found a zero-field mixed state at most temperatures (Fig. \ref{fig:loops}f), and we expect that the zero-field mixed state is present anywhere that the zero-field magnetic skyrmion state is available. 

These spin states are robust to variations in magnetic field while the sample is held at a constant temperature. Applying a magnetic field lower than the saturation magnetization grows domains aligned parallel to the applied field and shrinks domains aligned anti-parallel. Returning the sample to zero magnetic field will return the domains to their original size and shape. In many previously studied skyrmionic systems, skyrmions are temporarily stabilized by a perpendicular magnetic field and will disappear when the field is removed \cite{muhlbauer_skyrmion_2009}. In contrast, the low-energy cycloidal state in FCGT cannot be transformed into a skyrmion state by applying a magnetic field. Here, the spin state is uniquely determined by the temperature-magnetic field path history and is otherwise robust to changes in magnetic field.

\subsection*{Magnetic saturation}
\begin{figure}
    \centering
    \includegraphics[width = \linewidth]{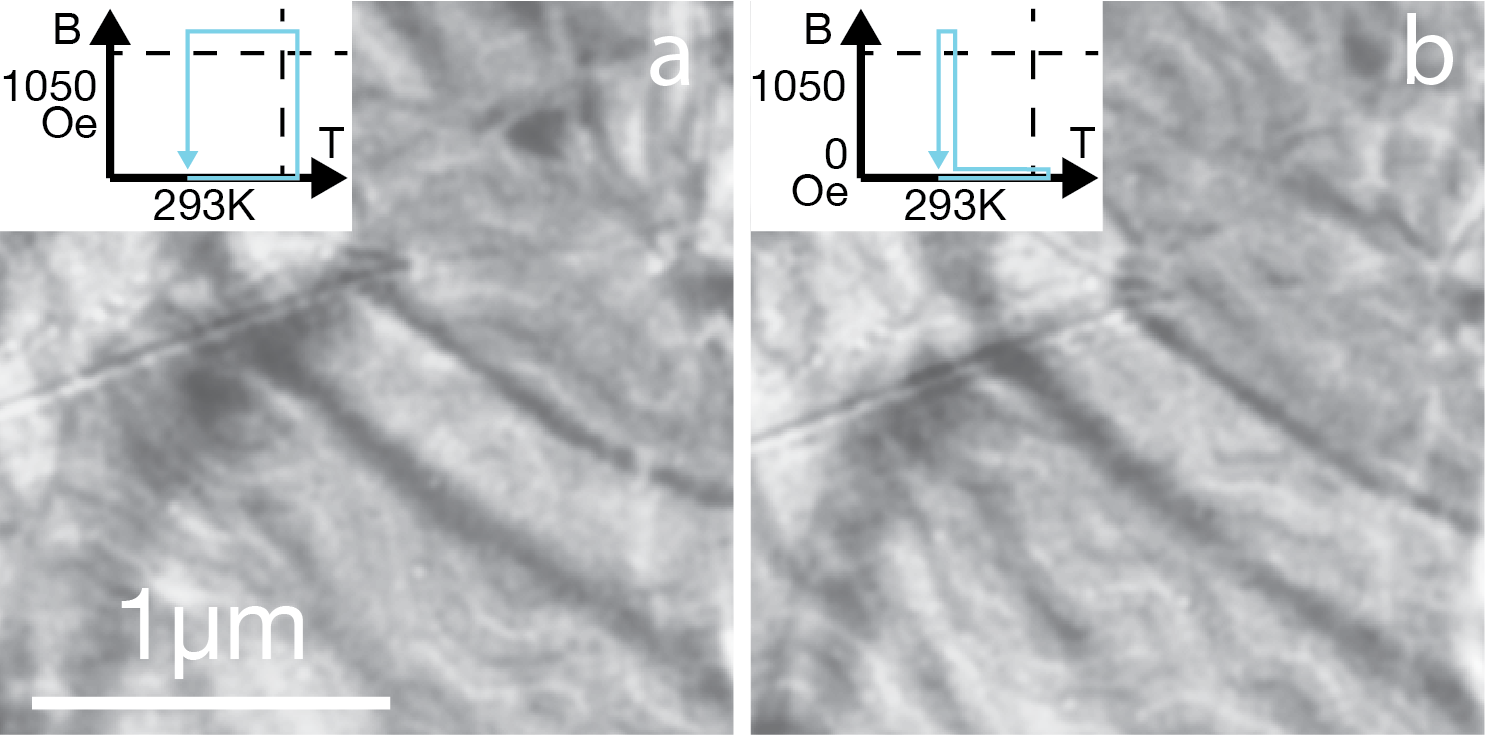}
    \caption{Magnetic saturation behavior of FCGT. a) LTEM image of a labyrinthine helical state resulting from field cooling at 1050 Oe (above the saturation magnetization) and 293K. b) A similar labyrinthine helical state resulting from isothermal magnetic saturation after zero-field cooling at 293K. Inset charts describe the paths through temperature-magnetic field phase space.}
    \label{fig:satBehavior}
\end{figure}

A magnetic field strong enough to saturate the material to a fully ferromagnetically aligned state will erase the temperature-magnetic field path history. Field-cooling with \Bc{} greater than the saturation field, B$_\textrm{sat}$, (inset in Fig. \ref{fig:satBehavior}a) yields a labyrinthine cycloidal state as in Figure \ref{fig:satBehavior}a. Applying a magnetic field greater than B$_\textrm{sat}$ at constant temperature to a zero-field-cooled cycloidal state at room temperature yields a similar labyrinthine cycloidal state (Fig. \ref{fig:satBehavior}b). Regardless of the temperature or field-cooling path traced beforehand, once we apply a magnetic field greater than B$_\textrm{sat}$,  the sample will be fully ferromagnetically aligned with the out-of-plane applied field. This erases the path history, so when the saturating field is removed the system will enter the labyrinthine cycloidal state. 

\subsection*{High and low temperatures}

\begin{figure}
    \centering
    \includegraphics[width = \linewidth]{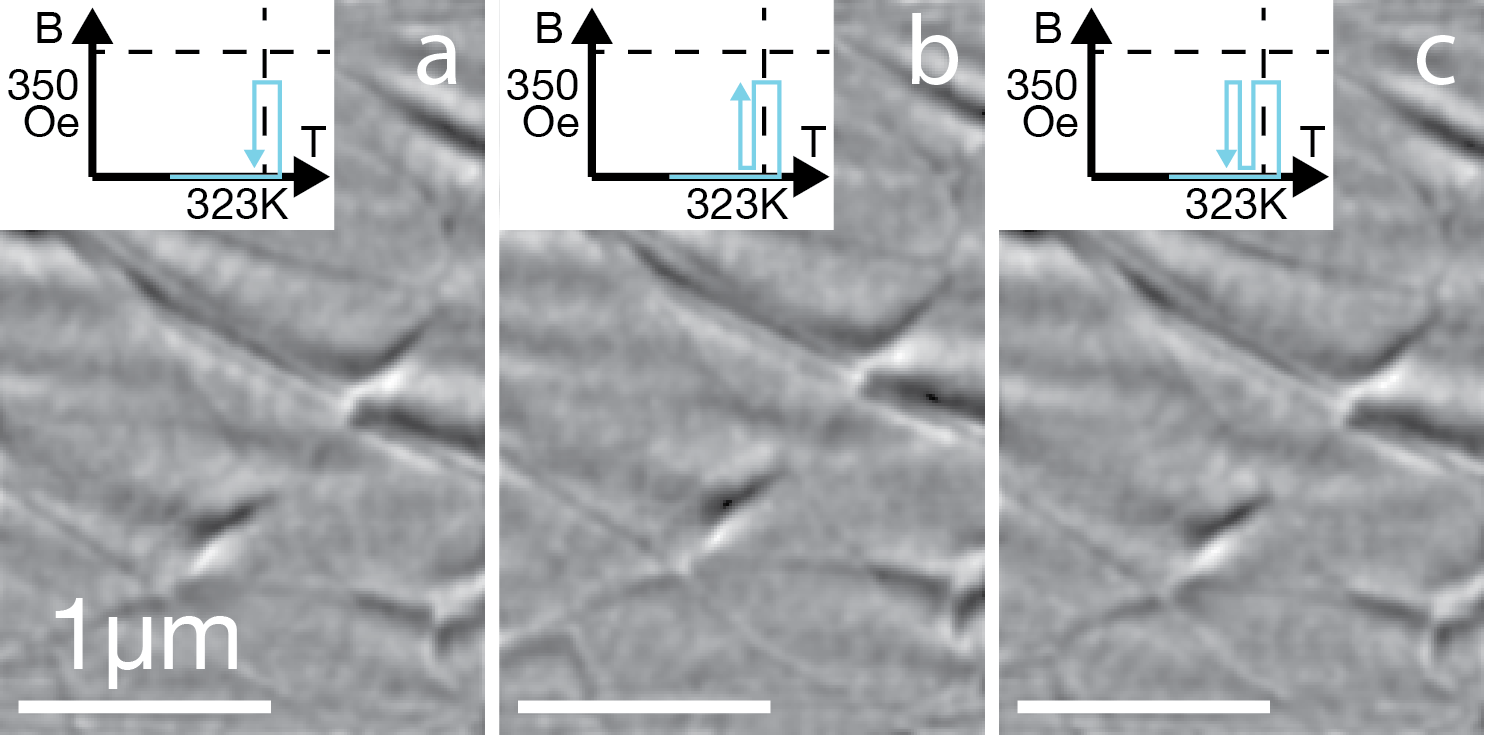}
    \caption{FCGT cycled in magnetic field at 323K. a) Initial zero-magnetic-field LTEM image showing helical domains after field cooling from above \Tc{} with \Bc{} = 350 Oe. b) LTEM image of magnetic skyrmions stabilized by a 350 Oe perpendicular magnetic field. c) Zero-field LTEM image of helical domains that reappeared after removing the perpendicular magnetic field.}
    \label{fig:highT}
\end{figure}

At high temperatures, metastable states are not available in FCGT. Field-cooling to 323K, approximately 10K below the Curie temperature, always results in the cycloidal state. 
In Figure \ref{fig:highT}a, we show a zero-field LTEM image of the cycloidal state selected by field-cooling at 350 Oe to 323K. The same field-cooling field yields a metastable zero-field skyrmion state at lower temperatures (Fig. \ref{fig:loops}c, d). When we apply a 350 Oe perpendicular magnetic field to the sample, it enters a stable skyrmion state (shown in LTEM with an applied field in Fig. \ref{fig:highT}b). When this magnetic field is removed, the system returns to the cycloidal state (Fig. \ref{fig:highT}c).  At high temperatures skyrmions are only stable under an applied magnetic field, similar to most reports of skyrmions in the literature \cite{nagaosa_topological_2013, muhlbauer_skyrmion_2009, yu_real-space_2010,yu_near_2011, ding_observation_2020, zhang_room-temperature_2022, montoya_tailoring_2017, wu_ne-type_2020}. At these high temperatures, the energy penalty to flip spins is small so the energy barrier between the skyrmion state and the cycloidal state is not large enough for zero field skyrmion metastability. 

\begin{figure}[t]
    \centering
    \includegraphics[width = \linewidth]{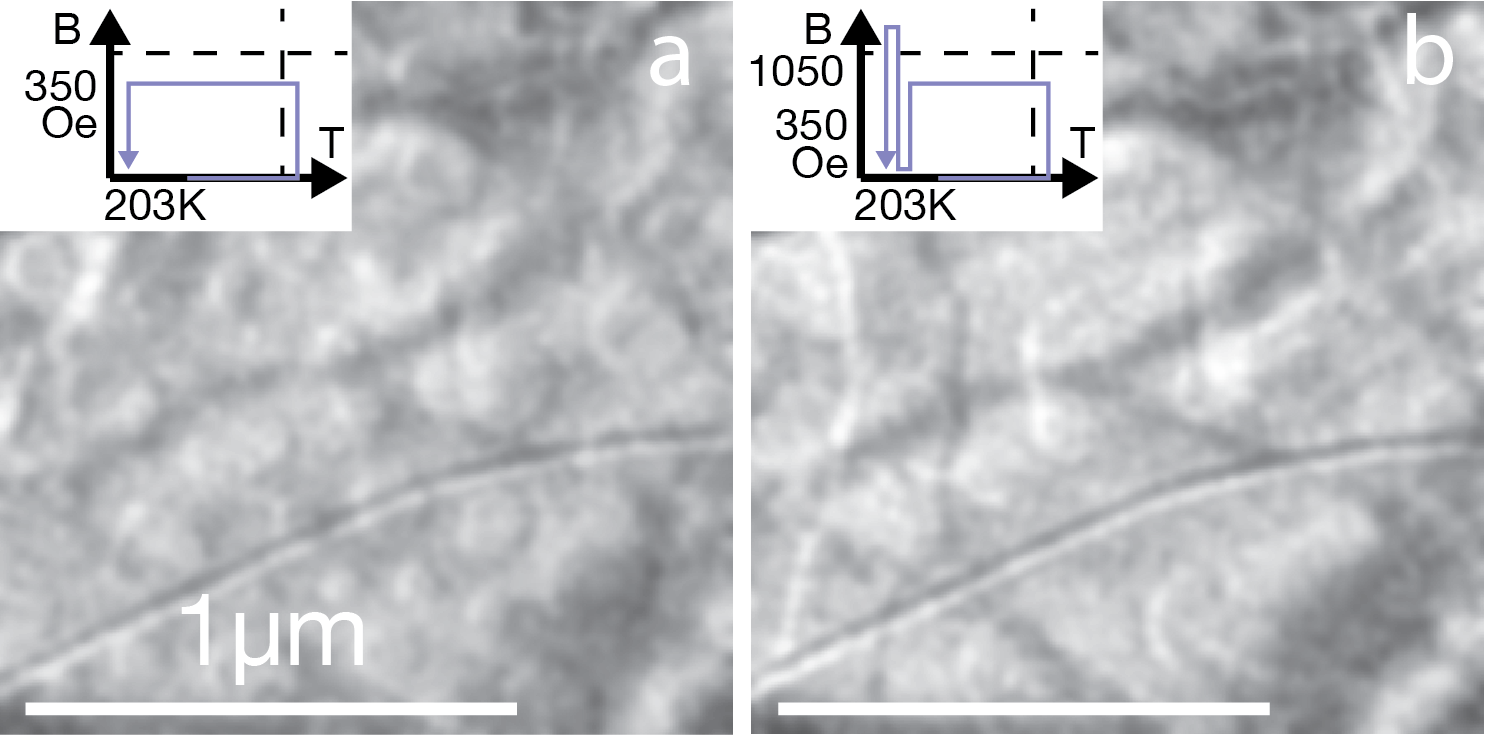}
    \caption{Low temperature magnetic textures in FCGT. a) LTEM image collected at 203K of circular domains in FCGT created by following the field cooling path at 350 Oe inset in (a). b) LTEM image of conventional N\'eel domains in FCGT created by applying then removing a saturating magnetic field to the texture in (a) (following the path inset in b). }
    \label{fig:lowTsat}
\end{figure}

At low temperatures (203K and below) the cycloidal state is not stable and the system adopts a classical domain structure with large, irregular magnetic domains and N\'eel domain walls. We field-cooled FCGT to 203K in a 350 Oe cooling field (path inset in Fig. \ref{fig:lowTsat}a). Due to the applied field during cooling, skyrmions nucleate once the sample reaches point ii on the cooling path (Fig. \ref{fig:loops}a). As the temperature decreases to 203K and the magnetic field is removed, these skyrmions expand into circular domains of irregular size as shown in Figure \ref{fig:lowTsat}a. We do not find the characteristic cycloidal state that we would expect at higher temperatures: after applying then removing a  magnetic field greater that B$_\textrm{sat}$ (path inset in Fig. \ref{fig:lowTsat}b), we find large, irregular N\'eel domains with no sign of cycloidal order (Fig. \ref{fig:lowTsat}b). This suggests that there is not helimagnetic ordering in FCGT at low temperatures---similar saturation experiments at higher temperatures consistently yielded a cycloidal magnetic state (Fig. \ref{fig:satBehavior}).

\subsection*{Comparison of magnetic skyrmions and bubbles}
\begin{figure}[ht]
    \centering
    \includegraphics[width = \linewidth]{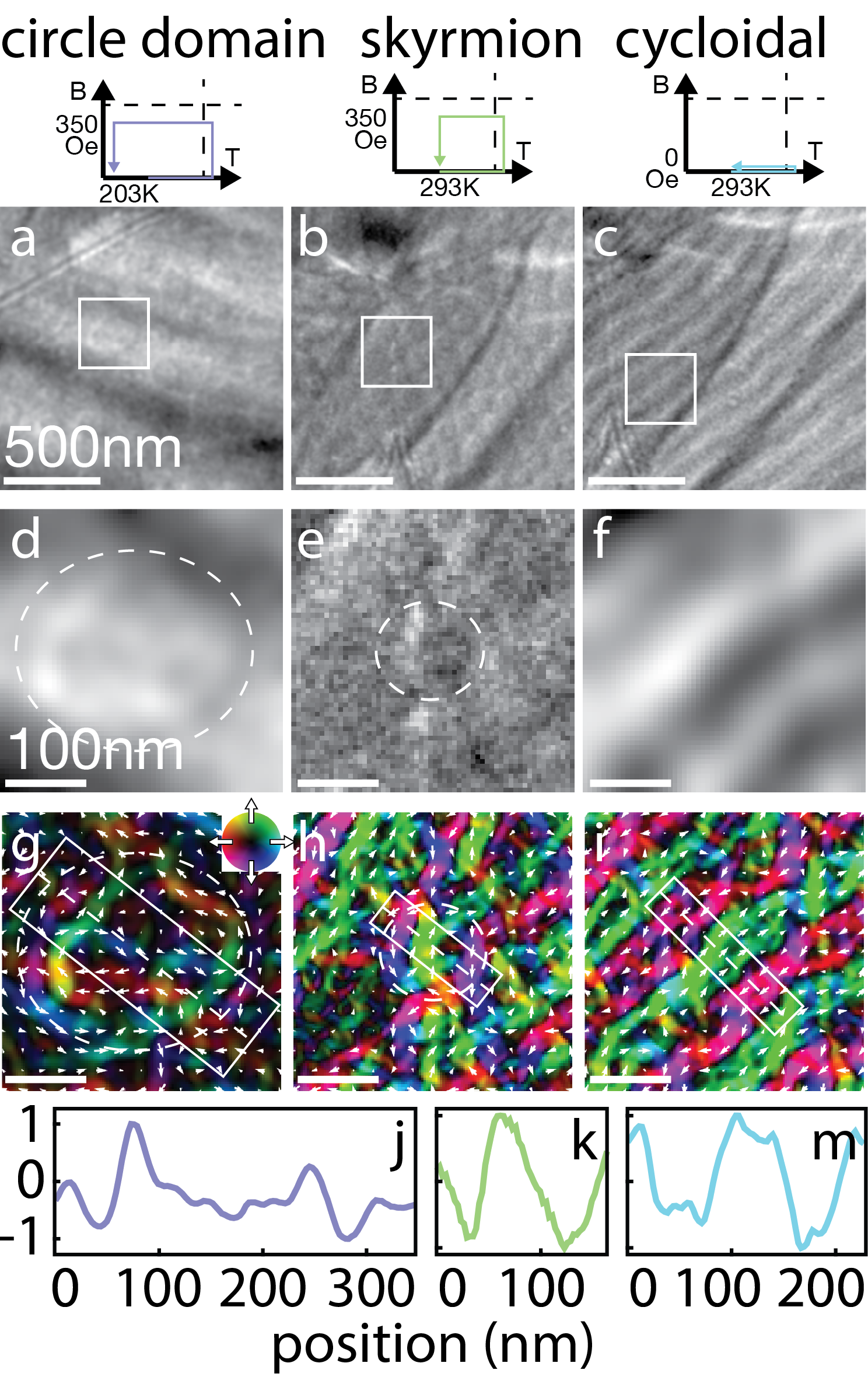}
    \caption{Comparison of circle domains and skyrmions. LTEM images of circle domains (a), skyrmions (b), and helical domains (c). Single circle domain (d), skyrmion (e), and helical stripe (f) selected from the white boxes in a, b, and c, respectively. The extent of the circle domain and the skyrmion are highlighted by a white dashed line. Magnetic induction maps of a circle domain (g), skyrmion (h), and helical stripe (i) reconstructed with a single-image TIE reconstruction from the LTEM images in d, e, and f. Magnetic induction direction is indicated by color shade, magnetic induction magnitude is indicated by color saturation, as in the inset colorwheel. Projection of in-plane magnetic induction vector across a circle domain (j), skyrmion (k), and helical stripe (m) from the diagonal dashed line in g, h, and i, respectively.}
    \label{fig:sky-circ}
\end{figure}

At low temperatures the helimagnetic ordering that stabilizes magnetic skyrmions and cycloids is not present. The irregularly-sized zero-field circle domains created with a 350 Oe cooling field at 203K (Fig. \ref{fig:sky-circ}a) are qualitatively distinct from the uniformly-sized zero-field skyrmions that we created at room temperature with a 350 Oe cooling field (Fig. \ref{fig:sky-circ}). By comparing the magnetic induction maps of each of these textures to that of the cycloidal state (Fig. \ref{fig:sky-circ}c), we can make a more quantitative comparison. We start by isolating a single circle domain (Fig. \ref{fig:sky-circ}d), skyrmion(Fig. \ref{fig:sky-circ}e), and cycloidal stripe (Fig. \ref{fig:sky-circ}f). To calculate the magnetic induction map of each magnetic texture we perform a set of single-image transport of intensity reconstructions using the open-source PyLorentz python package \cite{mccray_understanding_2021}. The transport of intensity equation (TIE) describes the translation of electron phase-shift to the measured detector intensity for thin samples in bright-field TEM \cite{gureyev_phase_1995}. These maps are shown in Figure \ref{fig:sky-circ}g-i. In each map the color corresponds to the in-plane direction of the vector field, and the saturation corresponds to the magnitude of the in-plane component of the vector field according to the inset color-wheel in Figure \ref{fig:sky-circ}g. We compare a projection of the components of the vector field parallel to the projection axis indicated in white for each magnetic texture, shown in Figures \ref{fig:sky-circ}j-m. The magnetic induction at the center of the circle domain is mostly out-of-plane, while the skyrmion and the cycloid have no extended area with out-of-plane magnetic induction. The LTEM contrast and magnetic induction of the skyrmion match previously reported N\'eel skyrmions \cite{pollard_observation_2017}, while the walls of the circle domains more closely match the contrast from classical N\'eel domains \cite{benitez_magnetic_2015}. Skyrmion size and shape depends on DMI-induced helimagnetic ordering that causes spin to continuously cant across a material---the growth of circle domains from skyrmions at low temperatures suggests that the helimagnetic ordering and the DMI are much weaker than the competing exchange interaction at 203K. 

\subsection*{Phase diagram of magnetic textures selected with varying field cooling schemes}
\begin{figure*}
    \includegraphics[width = \textwidth]{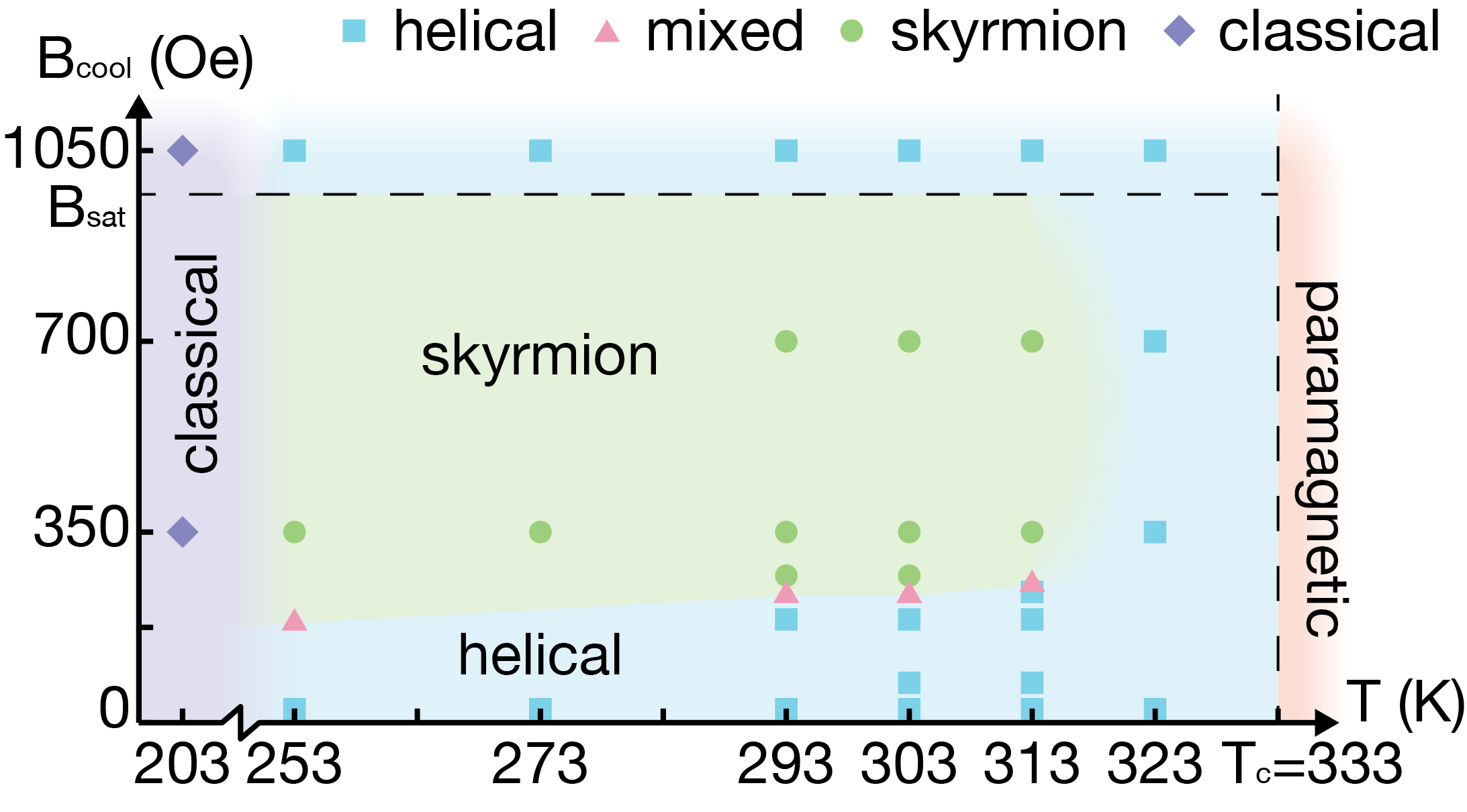}
    \caption{Phase diagram of magnetic textures selected with varying field cooling schemes. The shape and color of each point on the phase diagram corresponds to the magnetic texture imaged at zero magnetic field after a field cooling loop. For each point noted here, we heated the FCGT sample above \Tc{}, applied a magnetic field with magnitude indicated on the y-axis (\Bc{}), cooled the sample to the measurement temperature indicated on the x-axis, and removed the applied magnetic field. \label{fig:phases}}
\end{figure*}

We summarize the results of each field cooling experiment in a phase diagram with temperature along the x-axis and magnetic cooling field, \Bc{}, along the y-axis (Fig. \ref{fig:phases}). Every point on the phase diagram represents a single field-cooling experiment (data from each experiment is compiled in Supplemental Material Figs. S2-7) \cite{supp}. As an example, the point at 303K and 350 Oe represents the field cooling experiment during which we heated FCGT from 303K to above its \Tc{}, placed it in a 350 Oe magnetic cooling field, cooled it back down to 303K, and removed the magnetic field. After the field loop, a LTEM image is taken to determine which (meta)stable state the system is in at zero magnetic field. For the point at 350 Oe and 303K, zero-field skyrmions are metastable. We use the location of the mixed state (Fig. \ref{fig:states}g) as the boundary between the skyrmion and cycloidal state. We also know that any field cooling loop that ends above the Curie temperature (\Tc{}) will leave FCGT in the paramagnetic state, regardless of the magnetic cooling field applied. Magnetic saturation erases path history, so we did not collect full field cooling loops with \Bc{} $>$ B$_\textrm{sat}$ at every temperature---instead we measured the state after isothermal magnetic saturation, as this yields the same result as the full saturation-field-cooling loop. The result of each of these saturation experiments is a labyrinthine cycloidal state.

We can read this phase diagram as a map for processing FCGT to select a particular (meta)stable spin state. If we want to create skyrmions that are metastable at 313K, we choose any point on the map in the skyrmion region at 313K. Choosing the point at 700 Oe and 313K, we then heat FCGT to above \Tc{}, apply a perpendicular magnetic field of 700 Oe, cool it back to 313K, then remove the magnetic field. Now the FCGT should have metastable zero-field skyrmions.

\section*{Discussion}
Magnetic textures in FCGT behave differently under changing magnetic field than other classes of magnetic materials. In most skyrmionic materials, the skyrmion state is reached by applying a perpendicular magnetic field, which switches a cycloidal (or helical) spin state to a skyrmionic state \cite{muhlbauer_skyrmion_2009}. In those materials the magnetic texture is only determined by the temperature and applied magnetic field at the moment of measurement, so skyrmions are not stable at zero magnetic field.

In FCGT, the magnetic texture is determined by the temperature and magnetic field history. In this work we use the magnetic cooling field to select the magnetic ground state, then depending on the final temperature the system will get trapped in a metastable state or will relax to the ground state when the field is removed. Unlike skyrmions in other materials, the skyrmions in FCGT are metastable at room temperature and are robust to small changes in magnetic field.

The skyrmion state is metastable in FCGT at zero magnetic field because the system cannot overcome the energy barrier necessary to relax to the lower-energy cycloidal state. The cycloidal state is described by a spin cycloid in one direction throughout the material. The skyrmion state is described by three cycloids offset by 120$^\circ$ from one another \cite{muhlbauer_skyrmion_2009}. The topological stability of the skyrmion state arises because spins must flip 180$^\circ$ in several places throughout the material in order to switch from the skyrmion state to the cycloidal state and vice-versa. This topological stability is sufficient for skyrmions to be robust to locally smooth changes to magnetization via manipulation by external stimuli and interaction with material boundaries as long as the skyrmion state is the thermodynamically preferred spin arrangement \cite{jiang_blowing_2015, litzius_skyrmion_2017, yu_skyrmion_2012, zang_dynamics_2011}. But if the thermodynamic penalty for a discontinuous 180$^\circ$ spin flip is low, then the skyrmion state will relax to a cycloidal state as soon as the external magnetic field is removed. We observe this in FCGT only at high temperatures just below the Curie temperature (Fig. \ref{fig:highT}). Topological protection means that skyrmions are robust to smooth changes, but if there is no thermodynamic protection from discontinuous switches then the topological protection is not useful. At low temperatures, we observe that the topological protection of skyrmions persists regardless of the presence of the DMI. The circle domains observed in Figure \ref{fig:lowTsat} are topologically identical to skyrmions. Although the DMI is too weak at those temperatures to stabilize true skyrmions, the topological protection of the skymion spin arrangement prevented the circular structure from being destroyed. 

Metastable skyrmions created by field-cooling have been studied in MnSi, Co-Zn-Mn alloy, FeCoSi, and Cu$_2$OSeO$_3$ using small angle neutron scattering or topological hall effect measurements \cite{oike_interplay_2016, bauer_history_2016, karube_robust_2016, okamura_transition_2016}. This work shows similar phenomena in FCGT using a novel in-situ LTEM phase mapping technique for the first time. The LTEM images used in our technique reveal mixed and aperiodic magnetic states that could be missed or mistaken using other techniques.

\section*{Conclusion}
The spin state of a magnetic material is set by its path history---here we systematically followed paths through temperature-magnetic field phase space in order to map routes to the available spin states for FCGT. This mapping technique and these insights into spin state metastability are broadly applicable to the discovery of exotic spin textures in magnetic materials. As we study topologically protected spin states and how to manipulate them, it is crucial that we understand how to create a certain spin state and the conditions under which it will persist.

\section*{Acknowledgments}
This material is based upon work supported by the National Science Foundation Graduate Research Fellowship Program (R.Y.) under Grant No. DGE 1752814. Work at the University of California, Berkeley, the Molecular Foundry, and Lawrence Berkeley National Laboratory was funded by the U.S. DOE, Office of Science, Office of Basic Energy Sciences, Materials Sciences and Engineering Division under Contract No. DE-AC02-05CH11231 (Quantum Materials Program KC2202). Any opinions, findings, and conclusions or recommendations expressed in this material are those of the author(s) and do not necessarily reflect the views of the National Science Foundation.

\section*{Data Availability}
The data that support the findings of this article are openly available \cite{yalisove_data_2026}.

\bibliography{bibliography}

\end{document}


\title{Supplementary Material for\texorpdfstring{\\}{ }Mapping metastable magnetic textures in \texorpdfstring{\FCGT{}}{(Fe0.5Co0.5)5GeTe2} with in-situ Lorentz transmission electron microscopy}
\author{Reed Yalisove}
\affiliation{University of California, Berkeley, Department of Materials Science and Engineering, Berkeley, CA, United States}
\affiliation{Lawrence Berkeley National Lab, National Center for Electron Microscopy, Berkeley, CA, United States}

\author{Hongrui Zhang}
\affiliation{University of California, Berkeley, Department of Materials Science and Engineering, Berkeley, CA, United States}

\author{Xiang Chen}
\affiliation{University of California, Department of Physics, Berkeley, CA, United States}

\author{Fanhao Meng}
\author{Jie Yao}
\affiliation{University of California, Berkeley, Department of Materials Science and Engineering, Berkeley, CA, United States}
\affiliation{Lawrence Berkeley National Lab, Materials Sciences Division, Berkeley, CA, United States}

\author{Robert Birgeneau}
\affiliation{Lawrence Berkeley National Lab, Materials Sciences Division, Berkeley, CA, United States}
\affiliation{University of California, Department of Physics, Berkeley, CA, United States}

\author{Ramamoorthy Ramesh}
\affiliation{University of California, Berkeley, Department of Materials Science and Engineering, Berkeley, CA, United States}

\author{Mary C. Scott}
\affiliation{University of California, Berkeley, Department of Materials Science and Engineering, Berkeley, CA, United States}
\affiliation{Lawrence Berkeley National Lab, National Center for Electron Microscopy, Berkeley, CA, United States}

\maketitle

\section*{Weak skyrmion ordering}
\begin{figure}[ht]
\centering
\includegraphics[width = \textwidth]{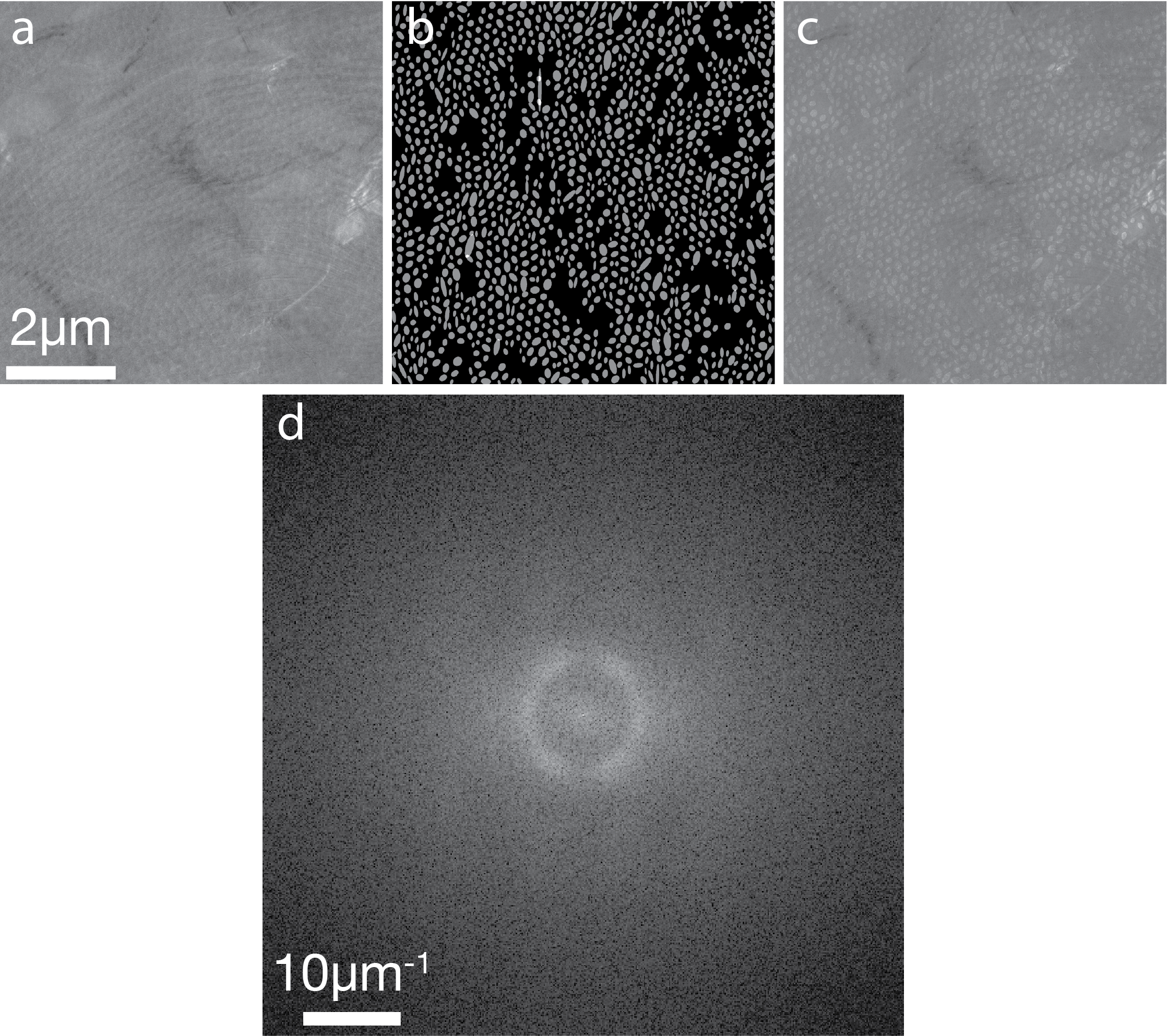}
\caption{Skyrmion ordering analysis. a) LTEM image of magnetic skyrmions taken after field-cooling at 350 Oe to 260K. A skyrmion detection algorithm was applied to this image to create the skyrmion mask in (b). c) Skyrmion mask overlaid on the original skyrmion image. d) Fast Fourier transform of binary skyrmion mask from (b), shown with log-scaled intensity}
\label{fig:ordering}
\end{figure}

Skyrmions in FCGT do not exhibit strong long-range ordering. In order to quantify skyrmion ordering, we developed an algorithm to reliably detect skyrmions in LTEM images. The N\'eel skyrmions in FCGT have low contrast and the image also includes strong diffraction contrast from bending and other defects. In order to isolate magnetic contrast, we use perform a transport of intensity equation (TIE) reconstruction on a focal stack of images collected in the same area as Figure \ref{fig:ordering}a using the PyLorentz python package \cite{mccray_understanding_2021}. With the resulting magnetic induction maps we use convolution-based template matching to identify skyrmions in the image. The matching template is based off of the expected magnetic induction of a N\'eel skyrmion with a predetermined estimated radius. We threshold and segment the resulting skyrmion mask, enforce minimum and maximum skyrmion area requirements to remove false identifications, then fit an ellipse to each identified skyrmion. The resulting skyrmion map is shown in Figure \ref{fig:ordering}b. In Figure \ref{fig:ordering}c the skyrmion map is overlaid on the original LTEM image to confirm the skyrmion identification. 

We analyze skyrmion ordering by taking a fast Fourier transform (FFT) of the skyrmion map, shown in log scale in Figure \ref{fig:ordering}d. This FFT has one prominent low-frequency ring, but no obvious peaks or second-order rings. This suggests that there is a consistent inter-skyrmion spacing, but no long-range ordering. 

\section*{Compilation of images used to create phase diagram}

\begin{figure}
	\centering
	\begin{subfigure}[t]{0.3\textwidth}
		\centering
		\includegraphics[width=\textwidth]{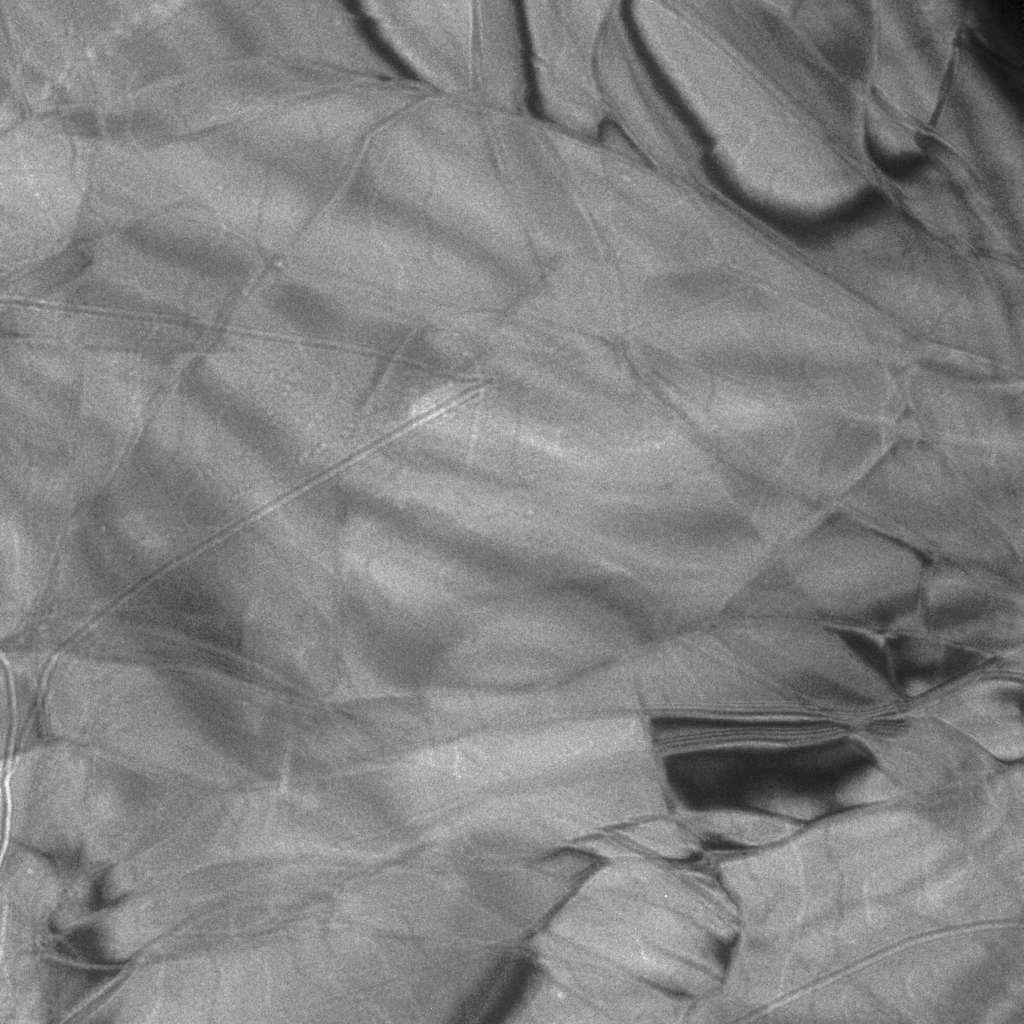}
		\caption{\Bc = 350 Oe, imaged after magnetic saturation}
	\end{subfigure}
	\begin{subfigure}[t]{0.3\textwidth}
		\centering
		\includegraphics[width=\textwidth]{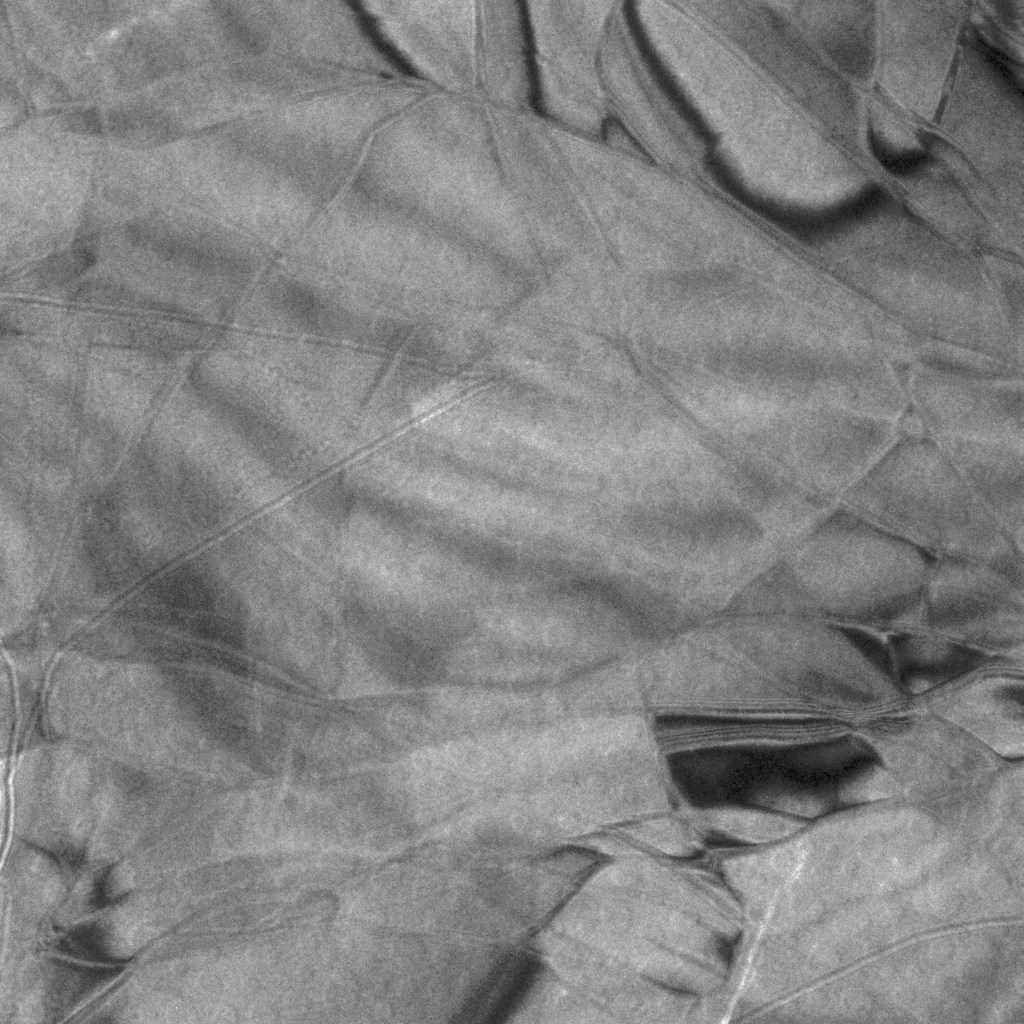}
		\caption{\Bc = 350 Oe}
	\end{subfigure}
	\caption{T = 203K, image field of view is 5.97$\mu$m}
	\label{fig: PDinfo200K}
\end{figure}

\begin{figure}
	\centering
	\begin{subfigure}[t]{0.3\textwidth}
		\centering
		\includegraphics[width=\textwidth]{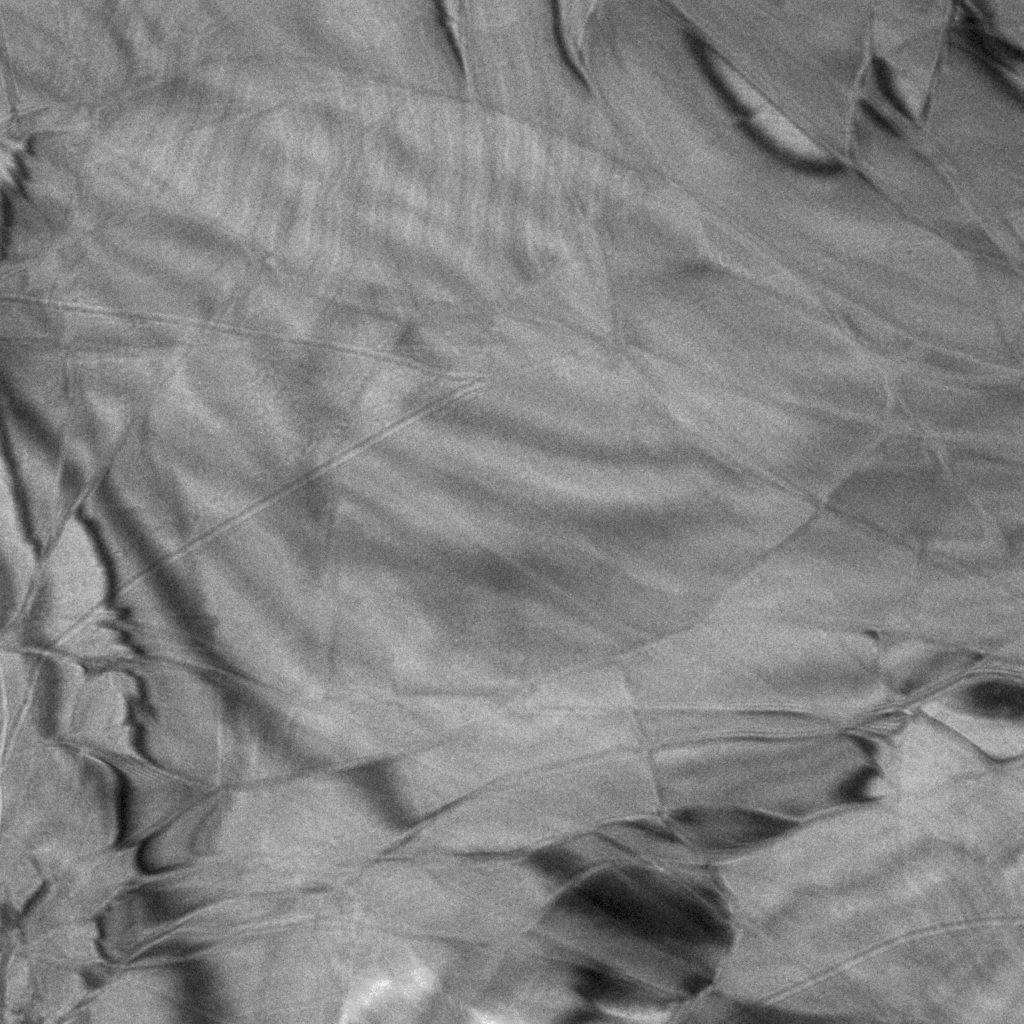}
		\caption{\Bc = 0 Oe}
	\end{subfigure}
	\begin{subfigure}[t]{0.3\textwidth}
		\centering
		\includegraphics[width=\textwidth]{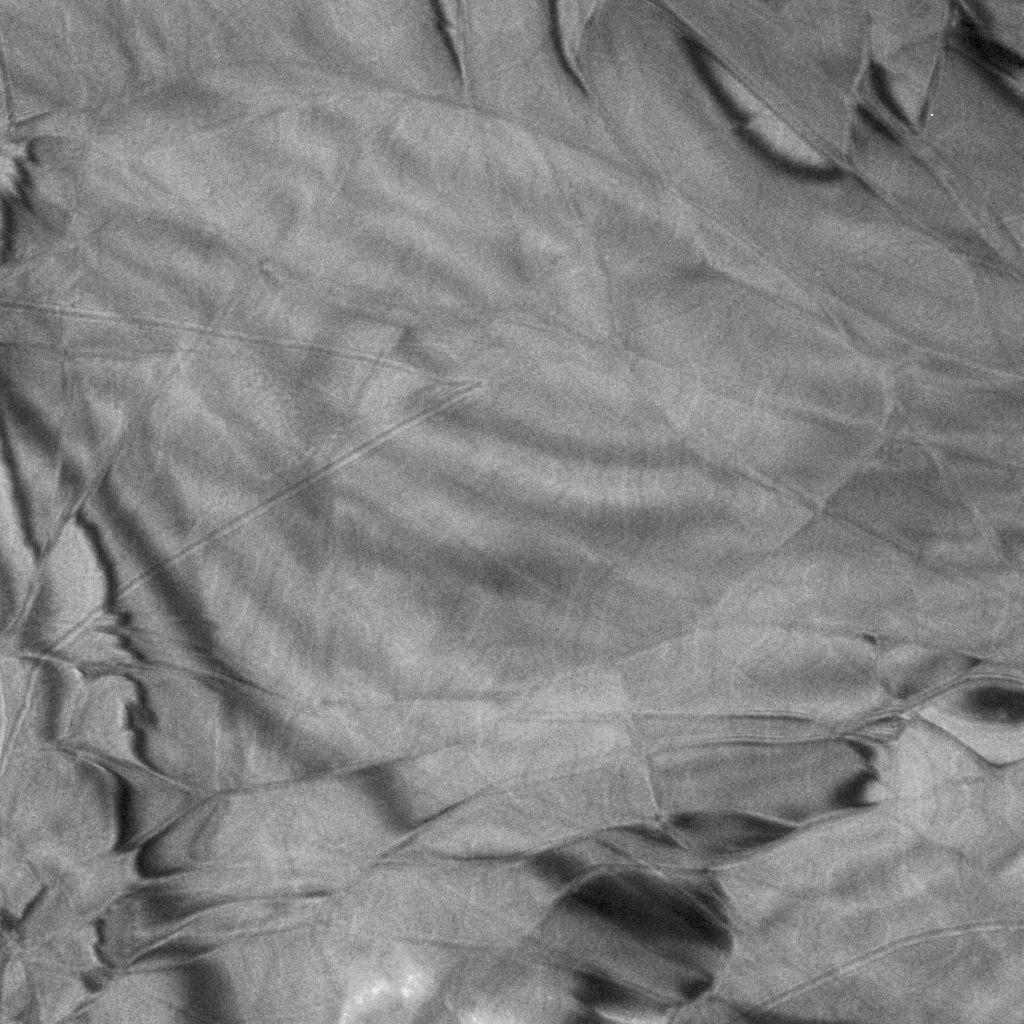}
		\caption{\Bc = 0 Oe, imaged after magnetic saturation}
	\end{subfigure}
	\begin{subfigure}[t]{0.3\textwidth}
		\centering
		\includegraphics[width=\textwidth]{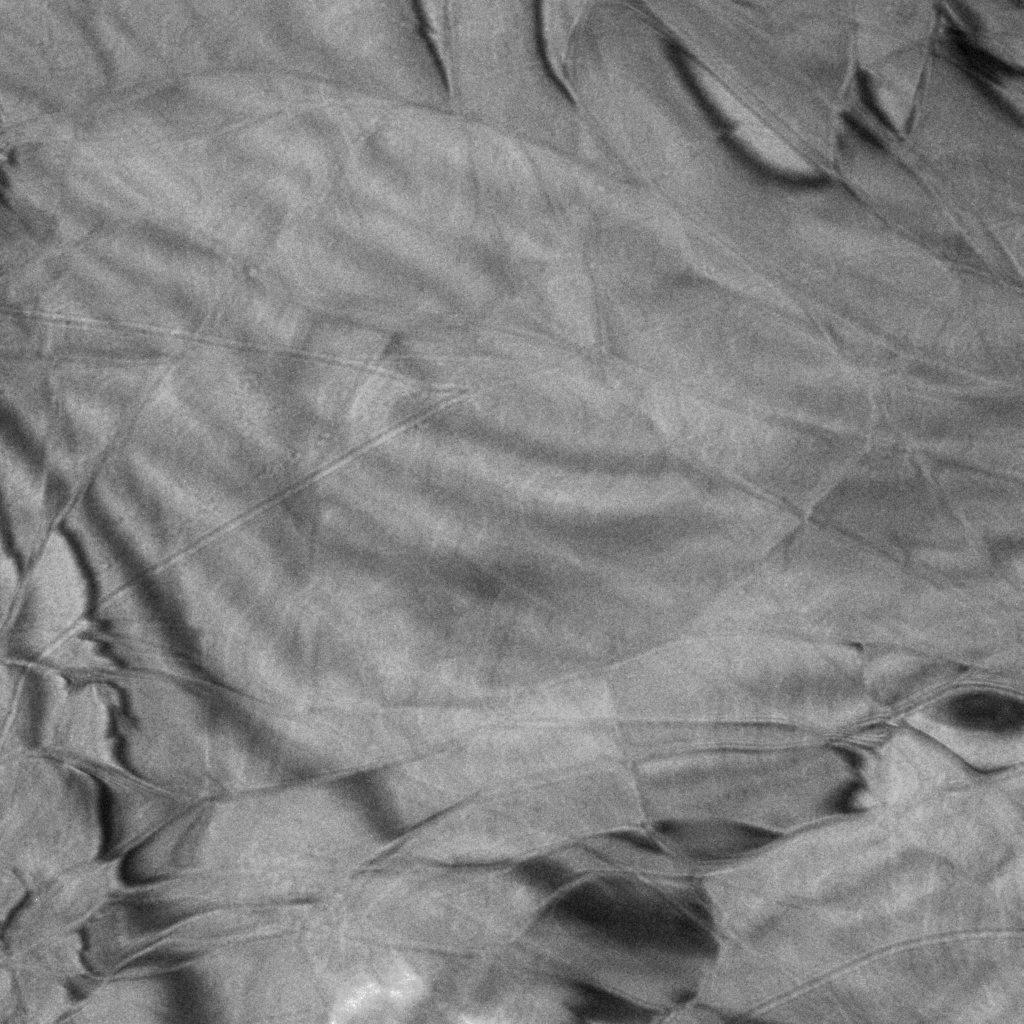}
		\caption{\Bc = 175 Oe}
	\end{subfigure}
	\begin{subfigure}[t]{0.3\textwidth}
		\centering
		\includegraphics[width=\textwidth]{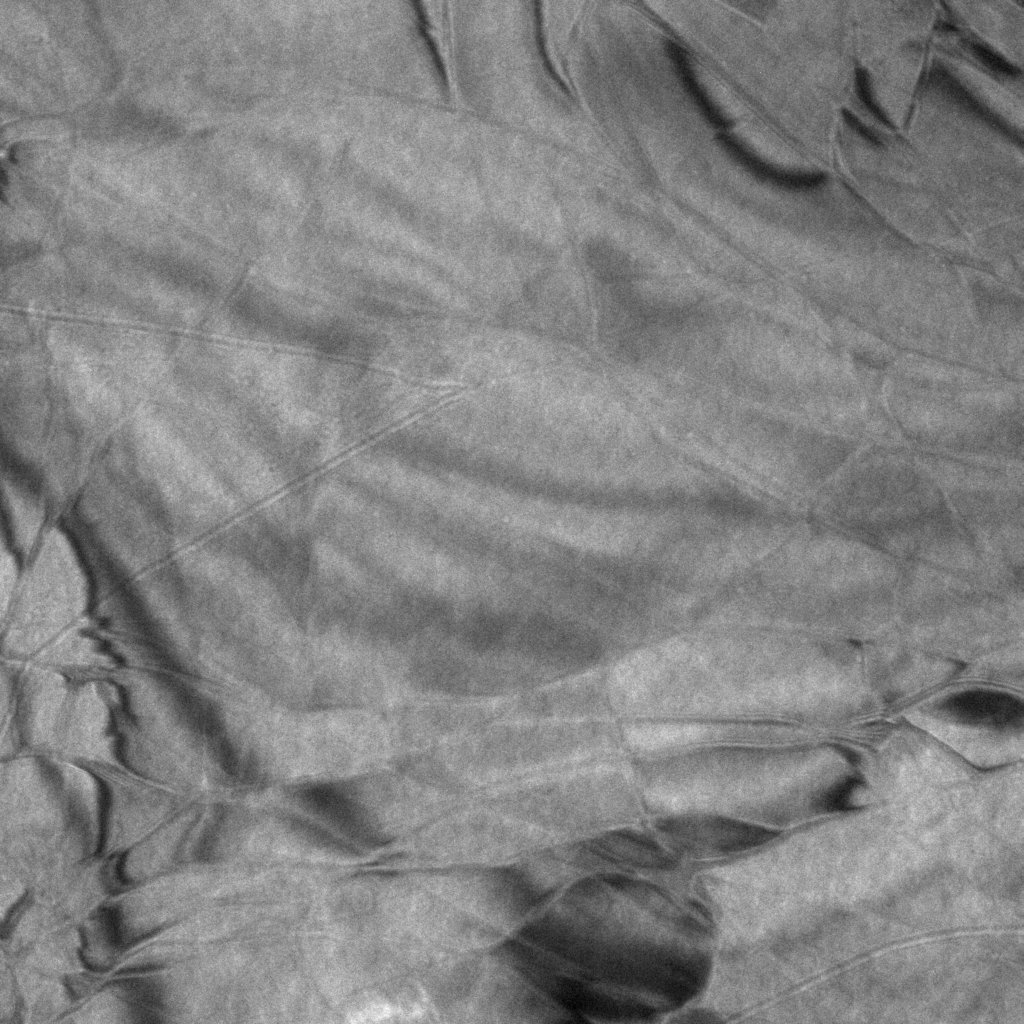}
		\caption{\Bc = 350 Oe}
	\end{subfigure}
	\caption{T = 253K, image field of view is 5.97$\mu$m}
	\label{fig: PDinfo250K}
\end{figure}

\begin{figure}
	\centering
	\begin{subfigure}[t]{0.3\textwidth}
		\centering
		\includegraphics[width=\textwidth]{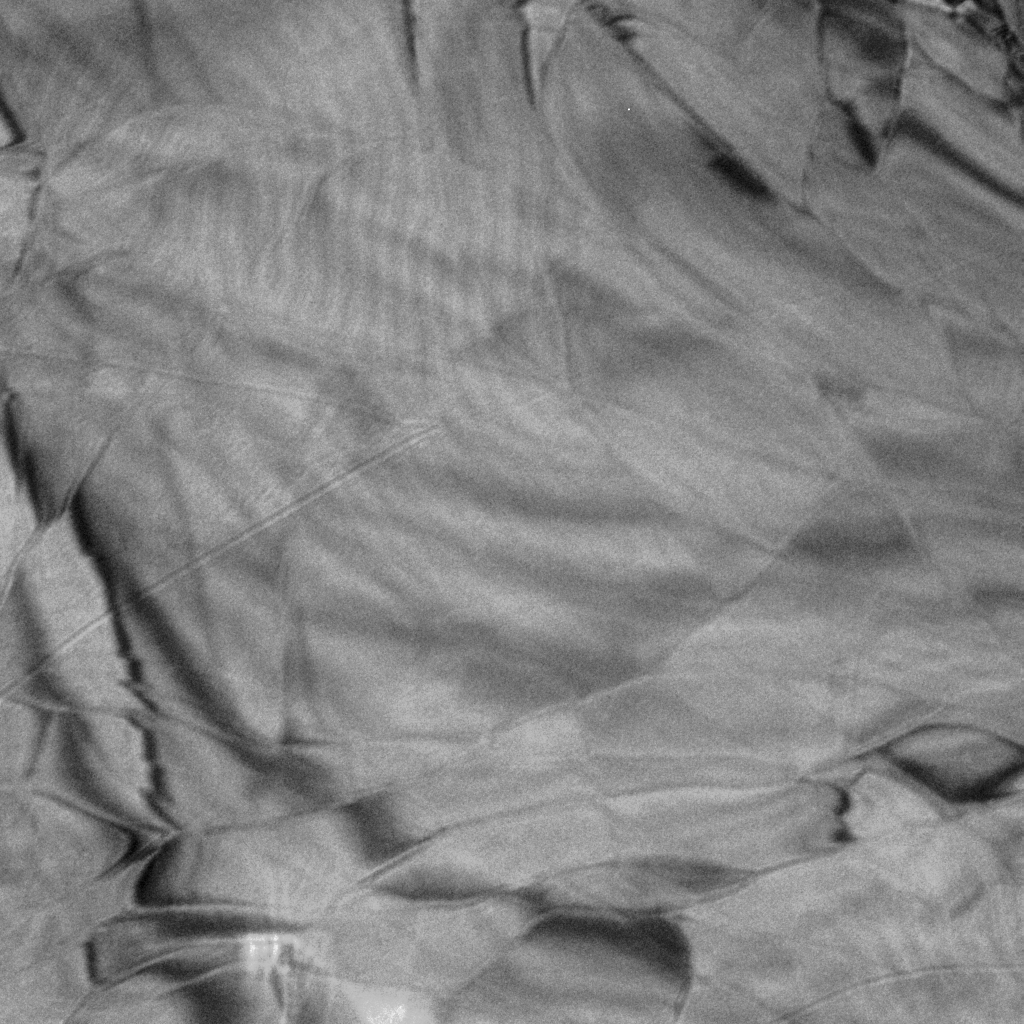}
		\caption{\Bc = 0 Oe}
	\end{subfigure}
	\begin{subfigure}[t]{0.3\textwidth}
		\centering
		\includegraphics[width=\textwidth]{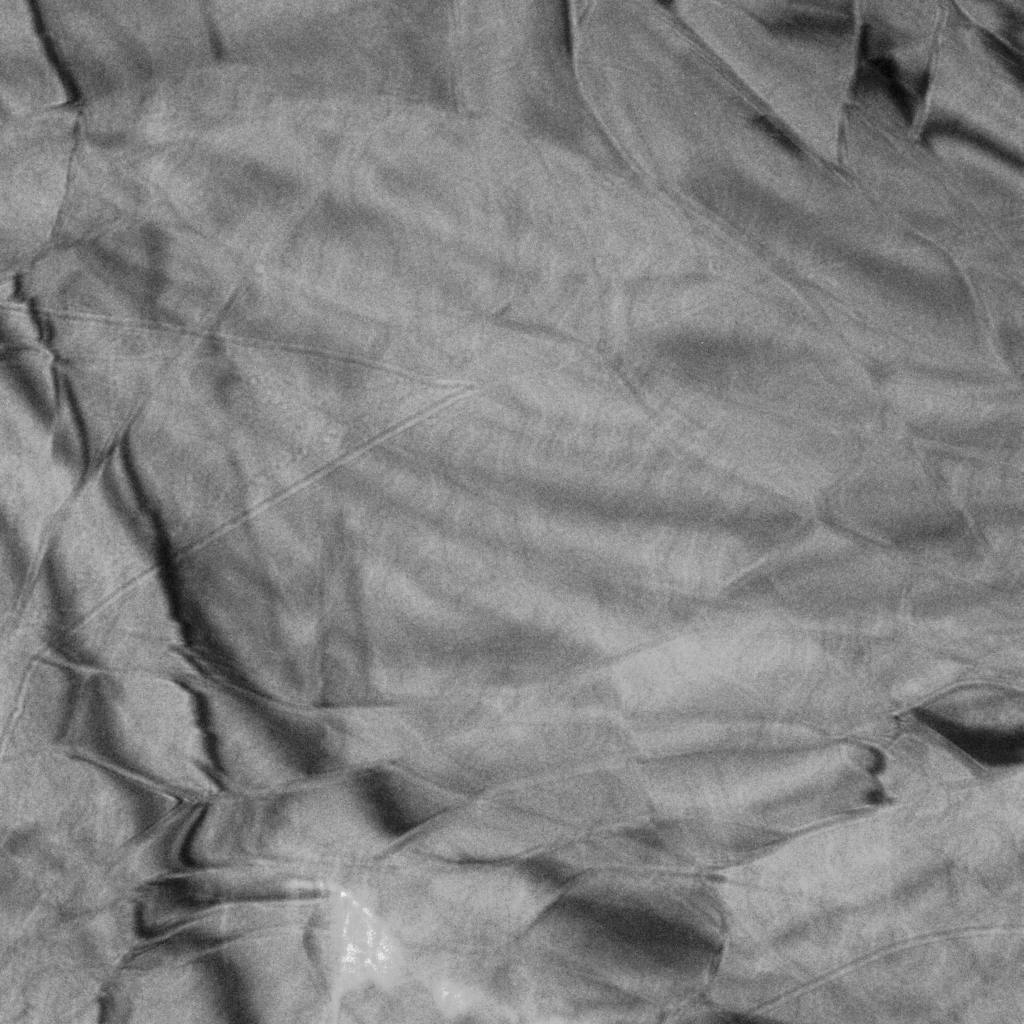}
		\caption{\Bc = 0 Oe, imaged after magnetic saturation}
	\end{subfigure}
	\begin{subfigure}[t]{0.3\textwidth}
		\centering
		\includegraphics[width=\textwidth]{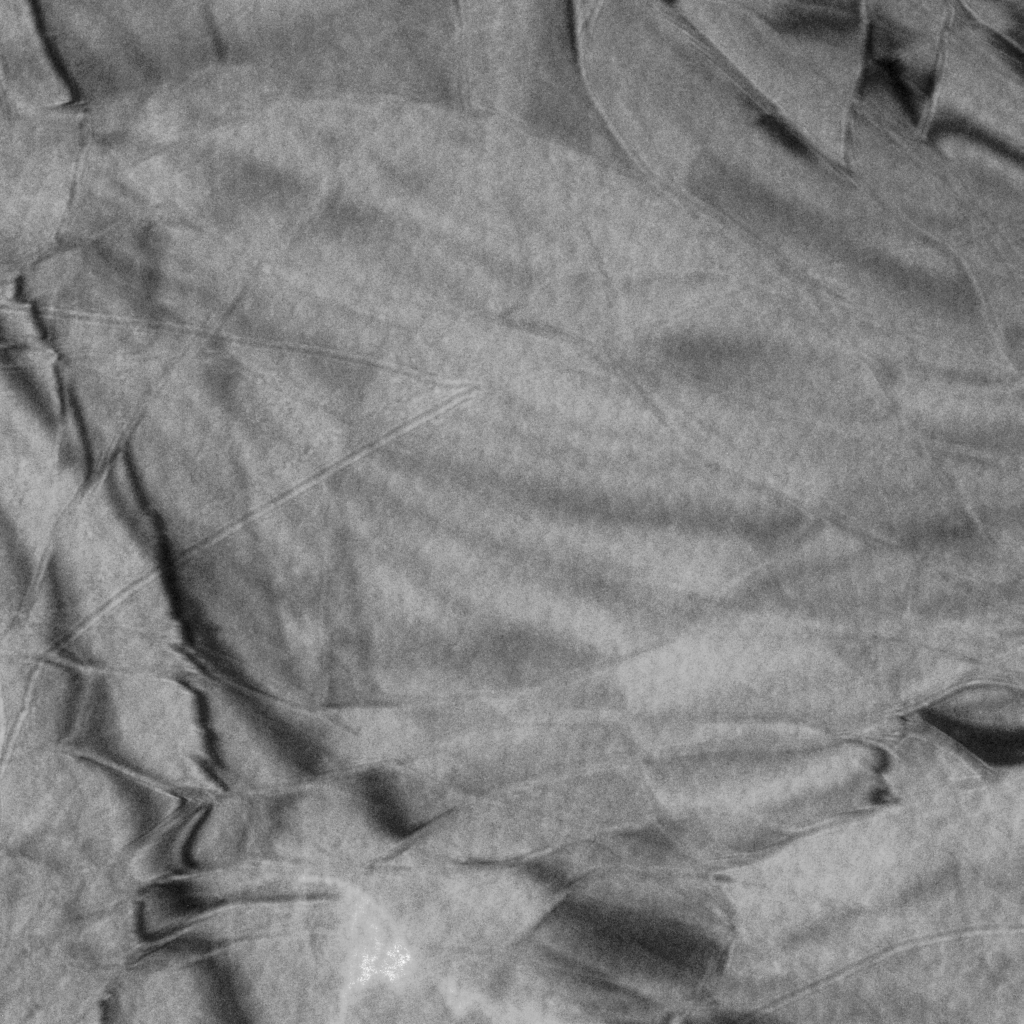}
		\caption{\Bc = 350 Oe}
	\end{subfigure}
	\caption{T = 273K, image field of view is 5.97$\mu$m}
	\label{fig: PDinfo273K}
\end{figure}

\begin{figure}
	\centering
	\begin{subfigure}[t]{0.3\textwidth}
		\centering
		\includegraphics[width=\textwidth]{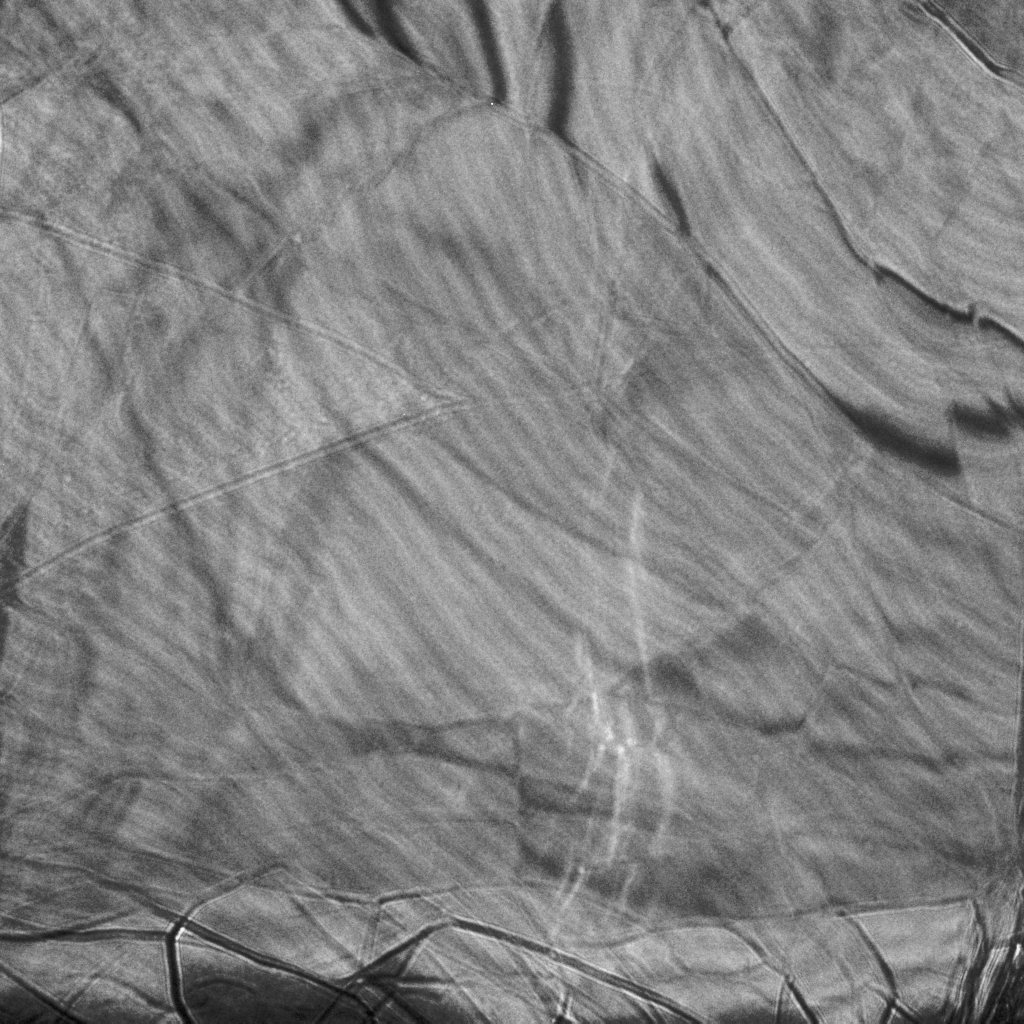}
		\caption{\Bc = 0 Oe}
	\end{subfigure}
	\begin{subfigure}[t]{0.3\textwidth}
		\centering
		\includegraphics[width=\textwidth]{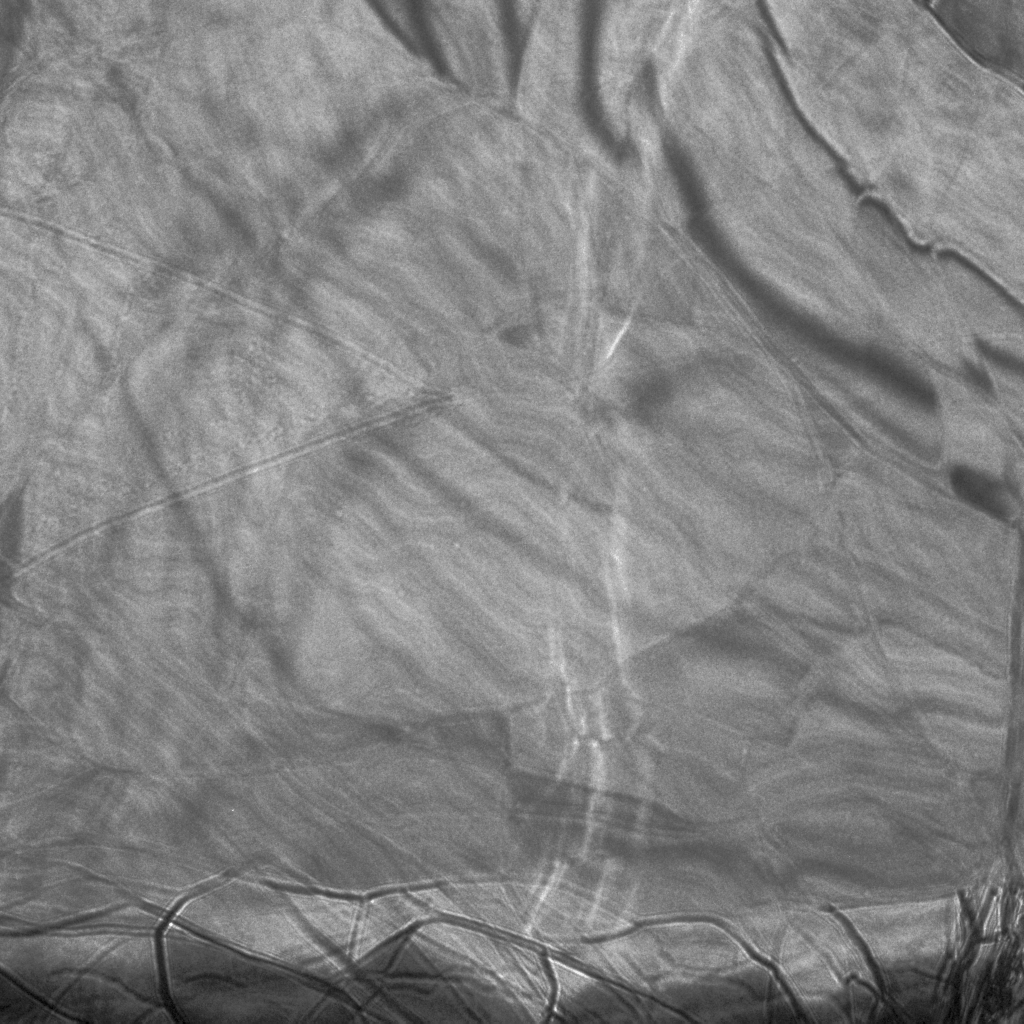}
		\caption{\Bc = 175 Oe}
	\end{subfigure}
	\begin{subfigure}[t]{0.3\textwidth}
		\centering
		\includegraphics[width=\textwidth]{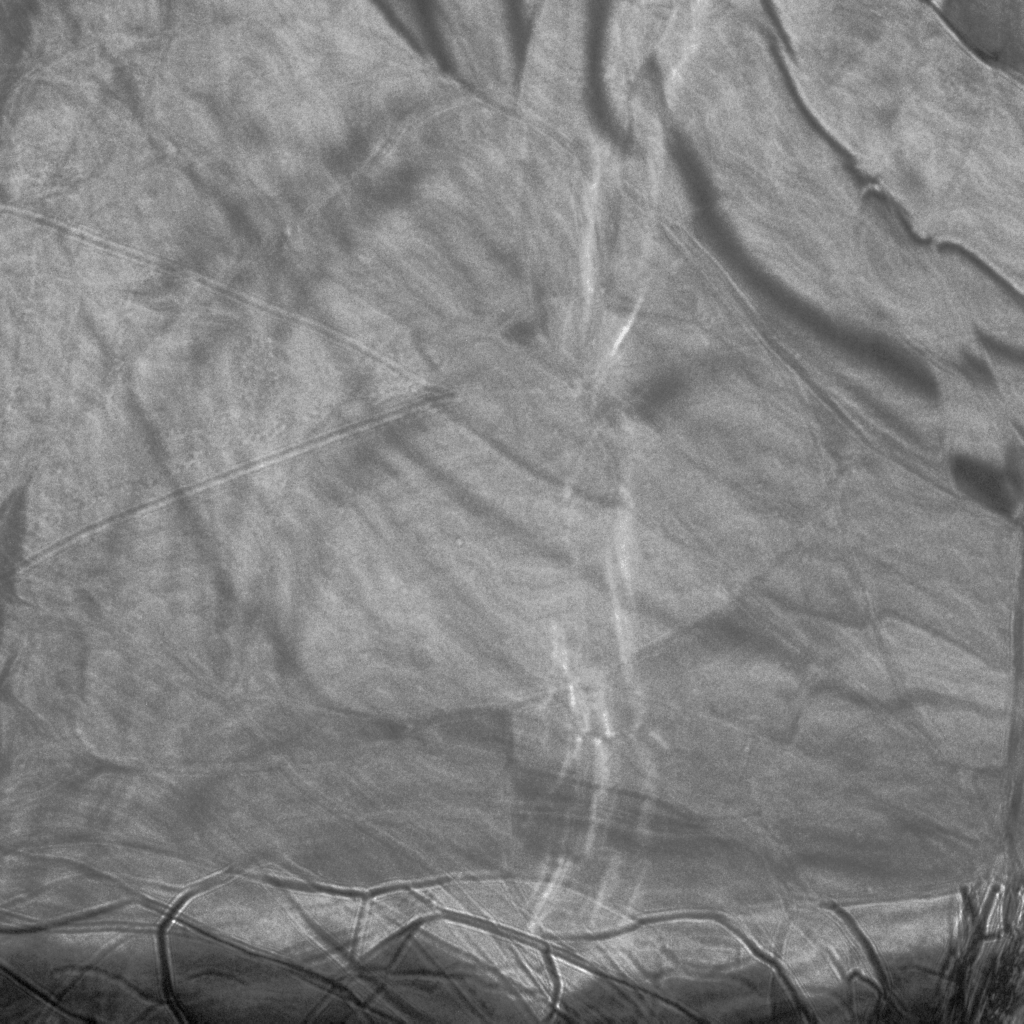}
		\caption{\Bc = 210 Oe}
	\end{subfigure}
	\begin{subfigure}[t]{0.3\textwidth}
		\centering
		\includegraphics[width=\textwidth]{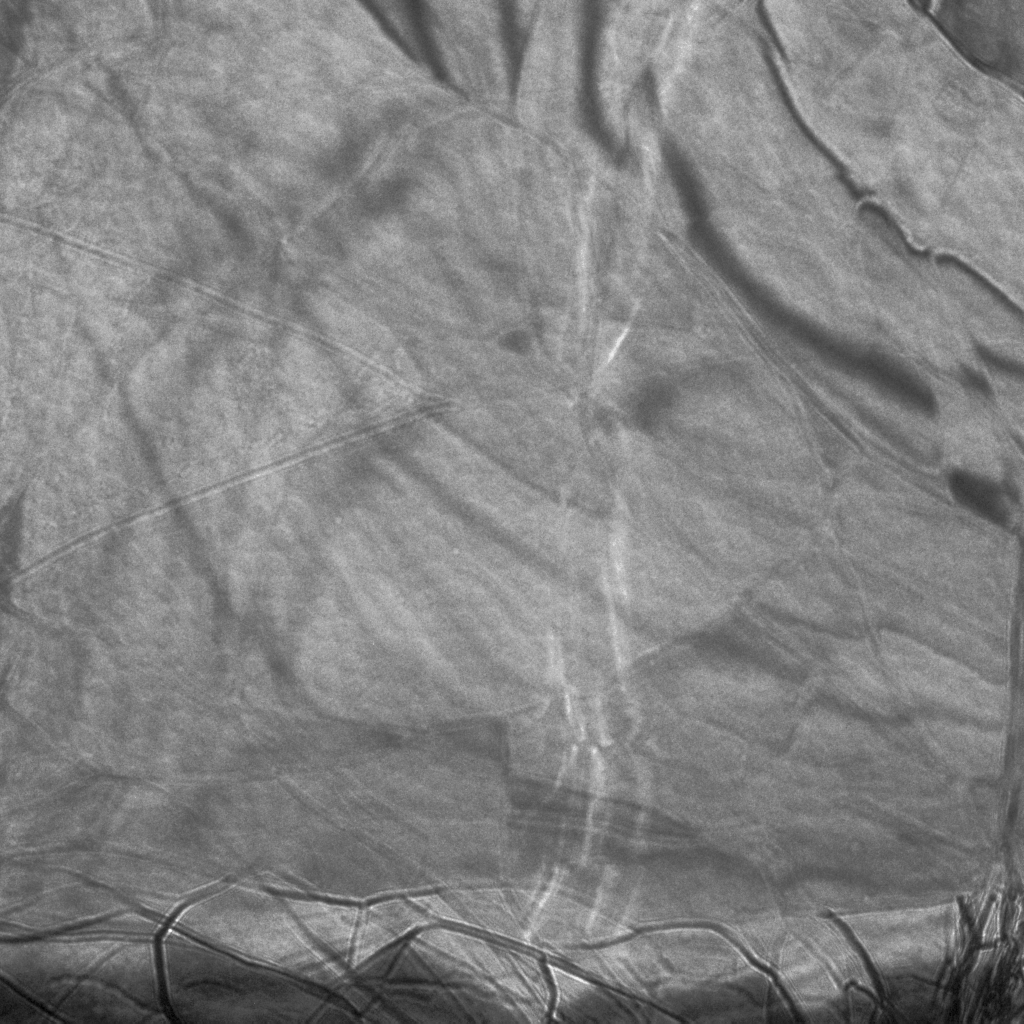}
		\caption{\Bc = 262 Oe}
	\end{subfigure}
	\begin{subfigure}[t]{0.3\textwidth}
		\centering
		\includegraphics[width=\textwidth]{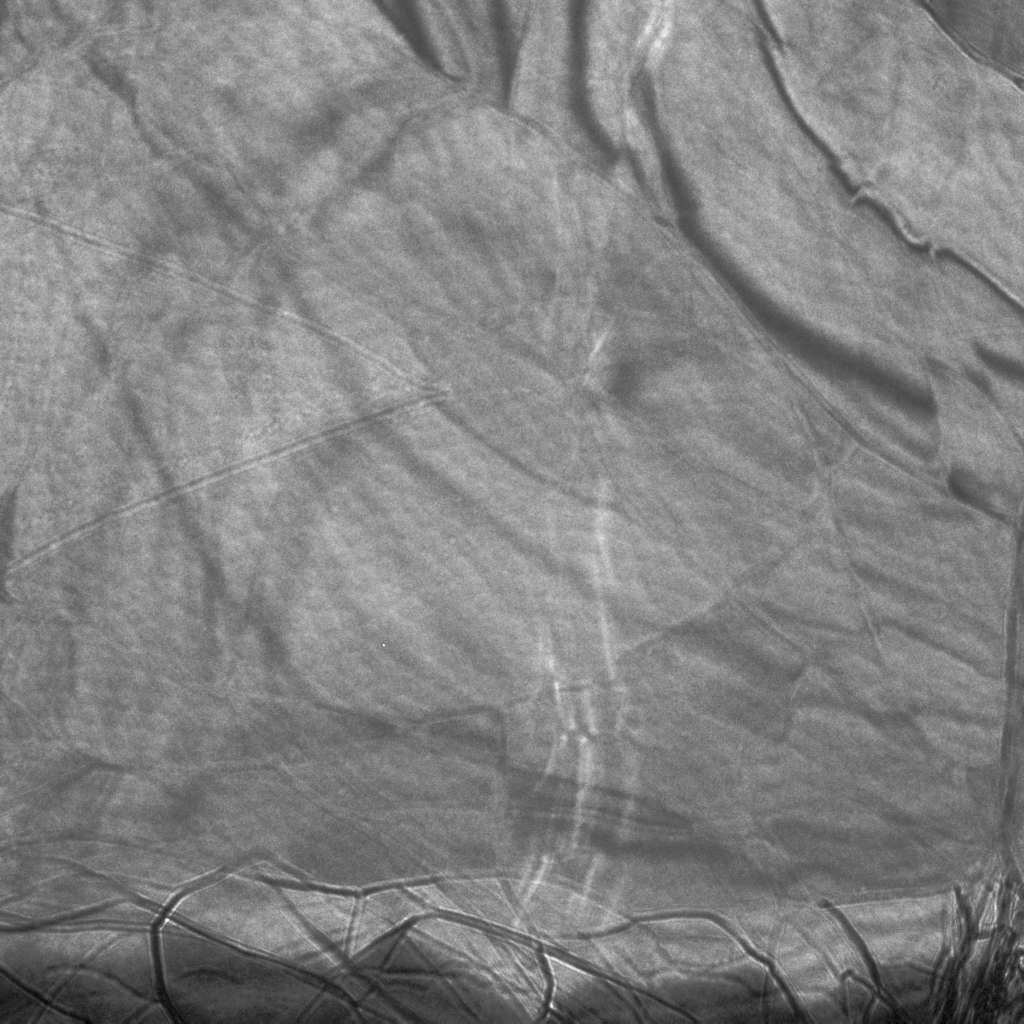}
		\caption{\Bc = 350 Oe}
	\end{subfigure}
	\begin{subfigure}[t]{0.3\textwidth}
		\centering
		\includegraphics[width=\textwidth]{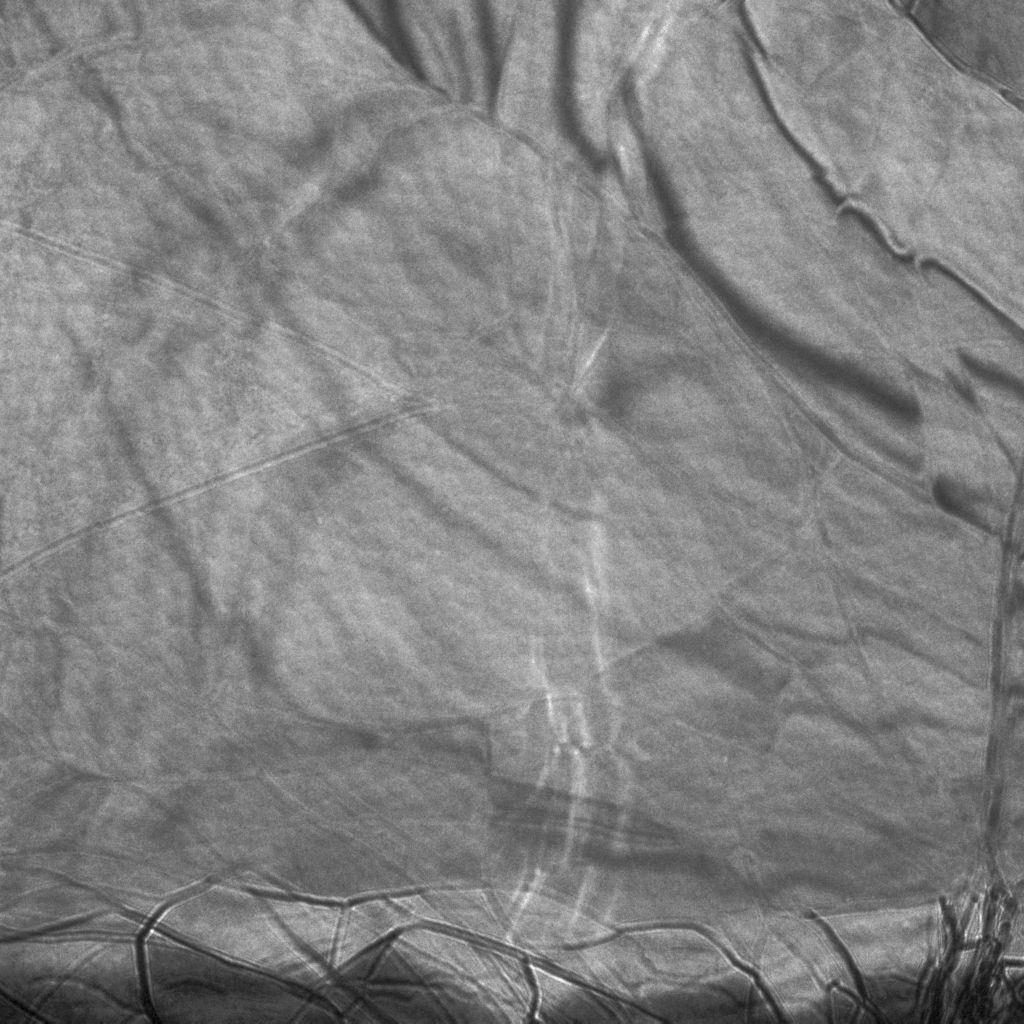}
		\caption{\Bc = 700 Oe}
	\end{subfigure}
	\begin{subfigure}[t]{0.3\textwidth}
		\centering
		\includegraphics[width=\textwidth]{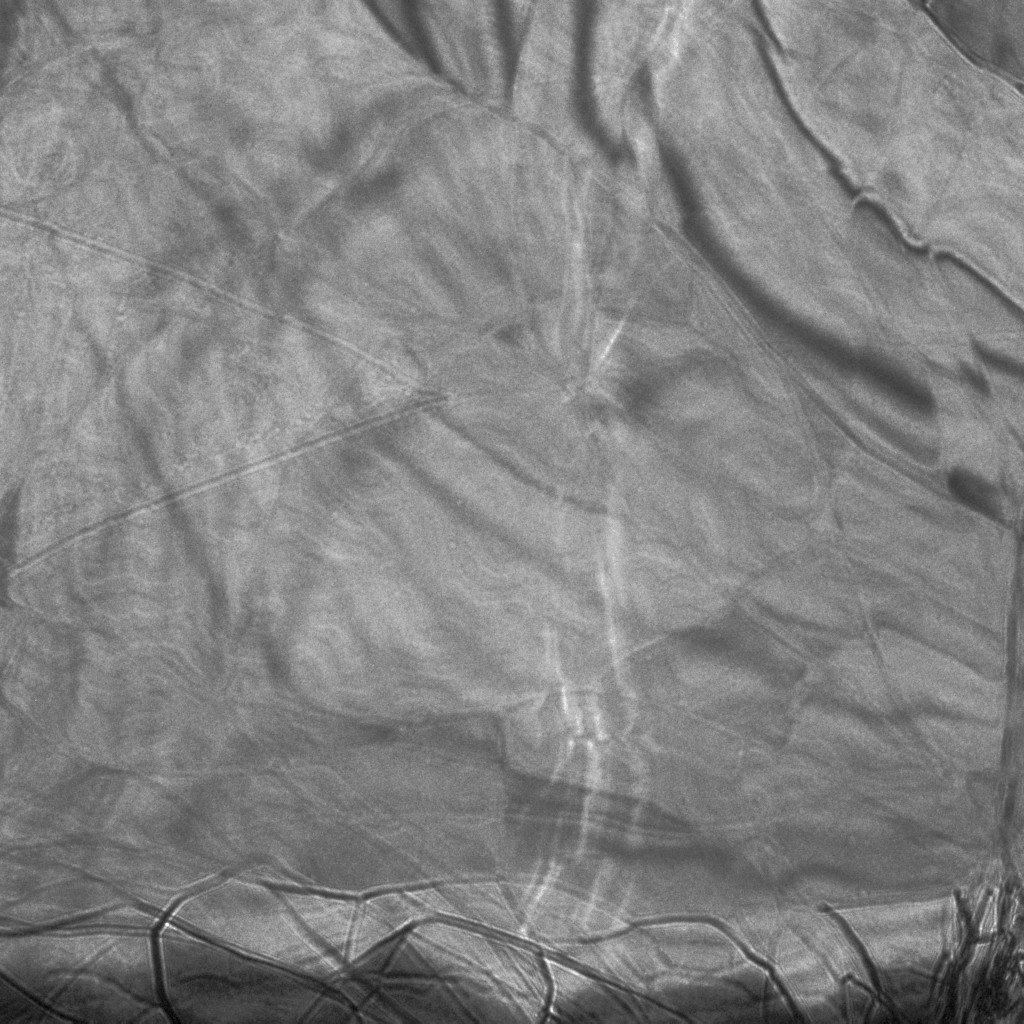}
		\caption{\Bc = 1050 Oe}
	\end{subfigure}
	\caption{T = 293K, image field of view is 5.97$\mu$m}
	\label{fig: PDinfo293K}
\end{figure}

\begin{figure}
	\centering
	\begin{subfigure}[t]{0.3\textwidth}
		\centering
		\includegraphics[width=\textwidth]{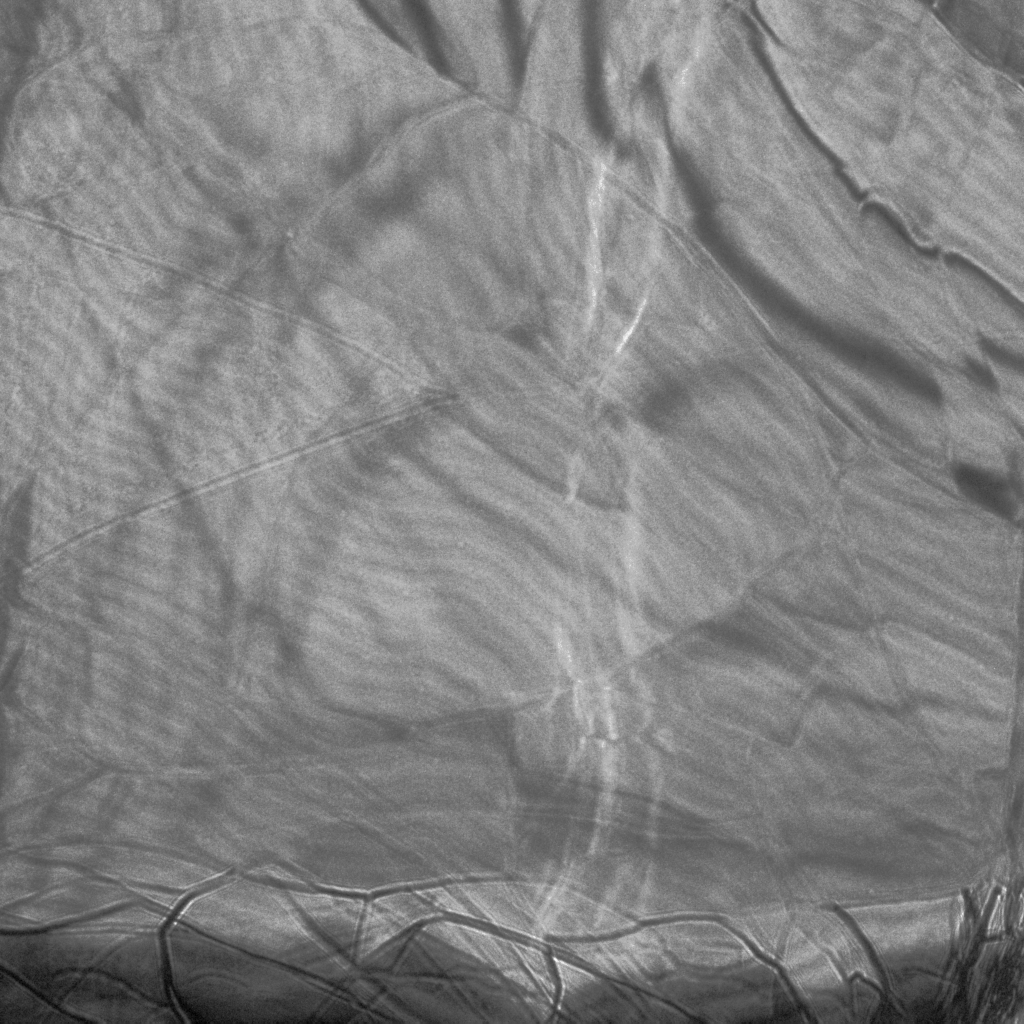}
		\caption{\Bc = 0 Oe}
	\end{subfigure}
	\begin{subfigure}[t]{0.3\textwidth}
		\centering
		\includegraphics[width=\textwidth]{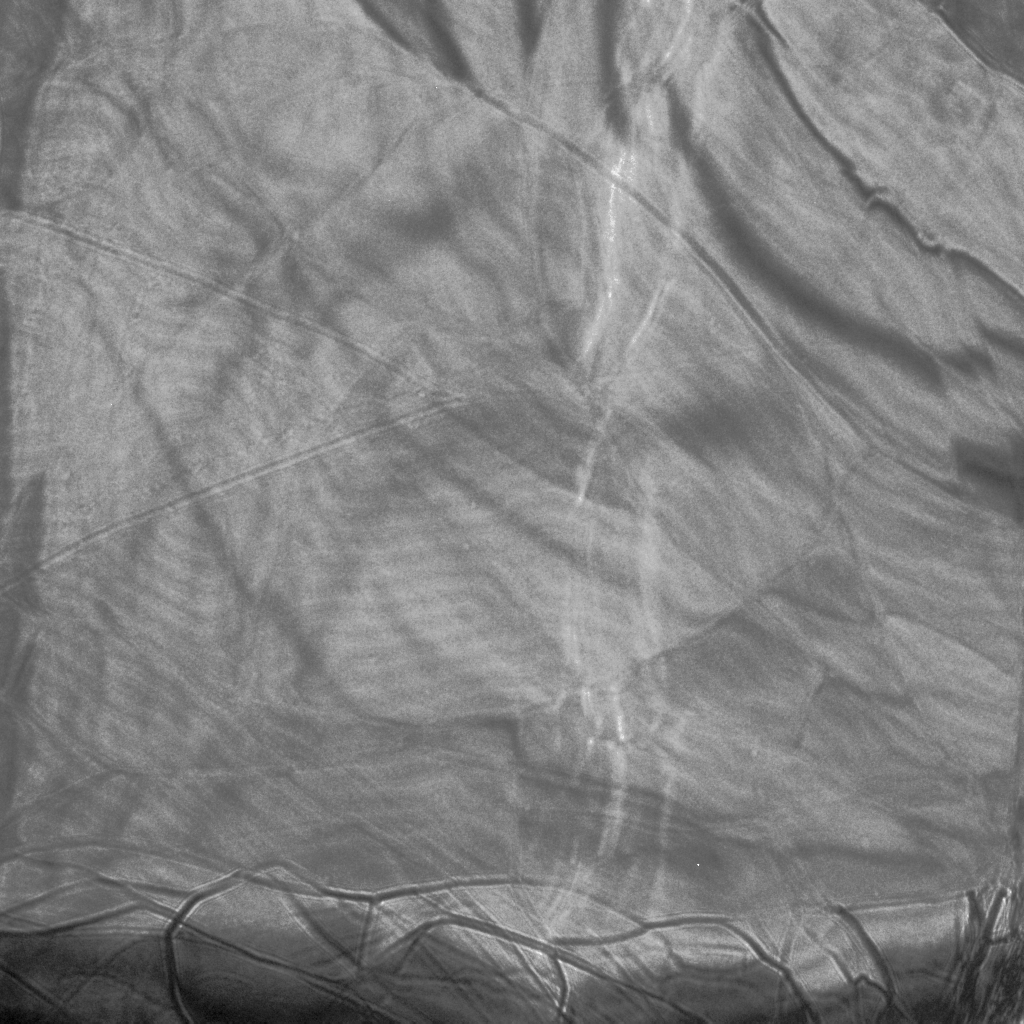}
		\caption{\Bc = 70 Oe}
	\end{subfigure}
	\begin{subfigure}[t]{0.3\textwidth}
		\centering
		\includegraphics[width=\textwidth]{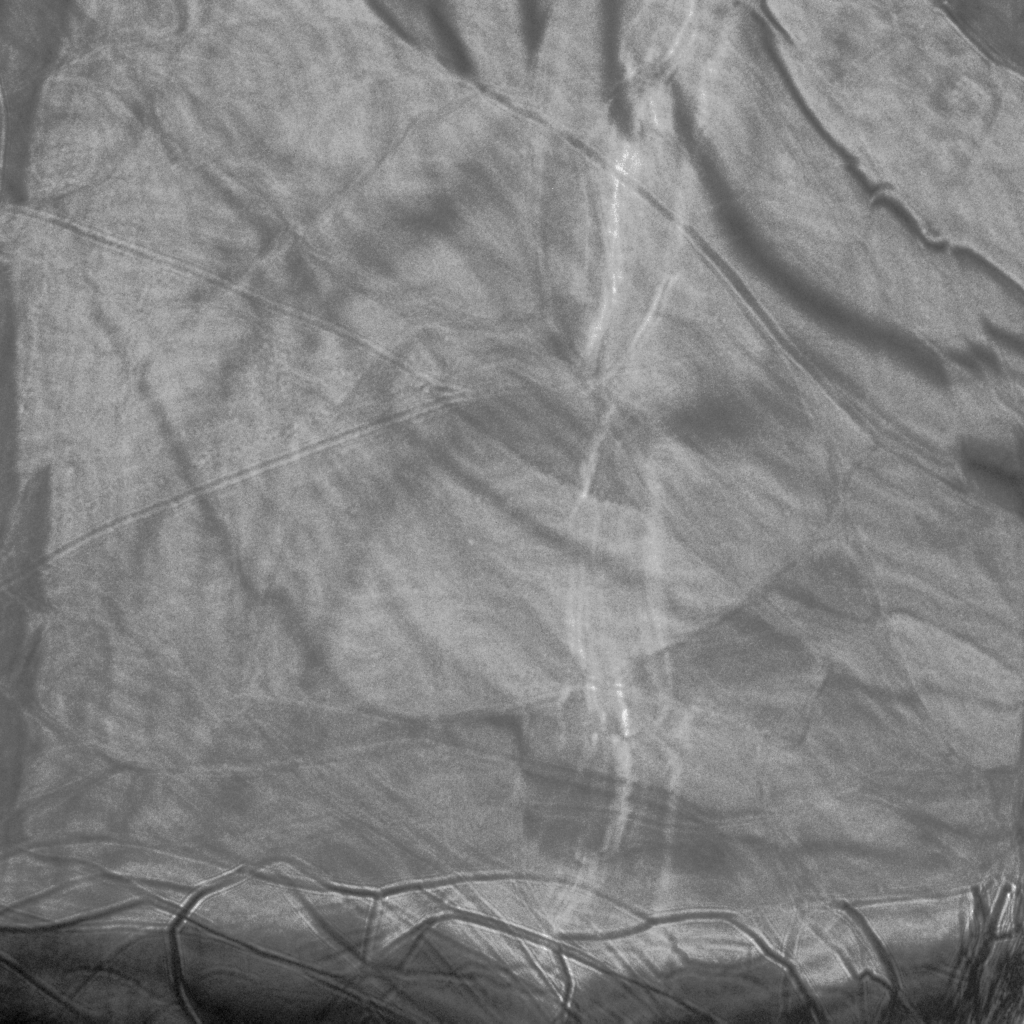}
		\caption{\Bc = 175 Oe}
	\end{subfigure}
	\begin{subfigure}[t]{0.3\textwidth}
		\centering
		\includegraphics[width=\textwidth]{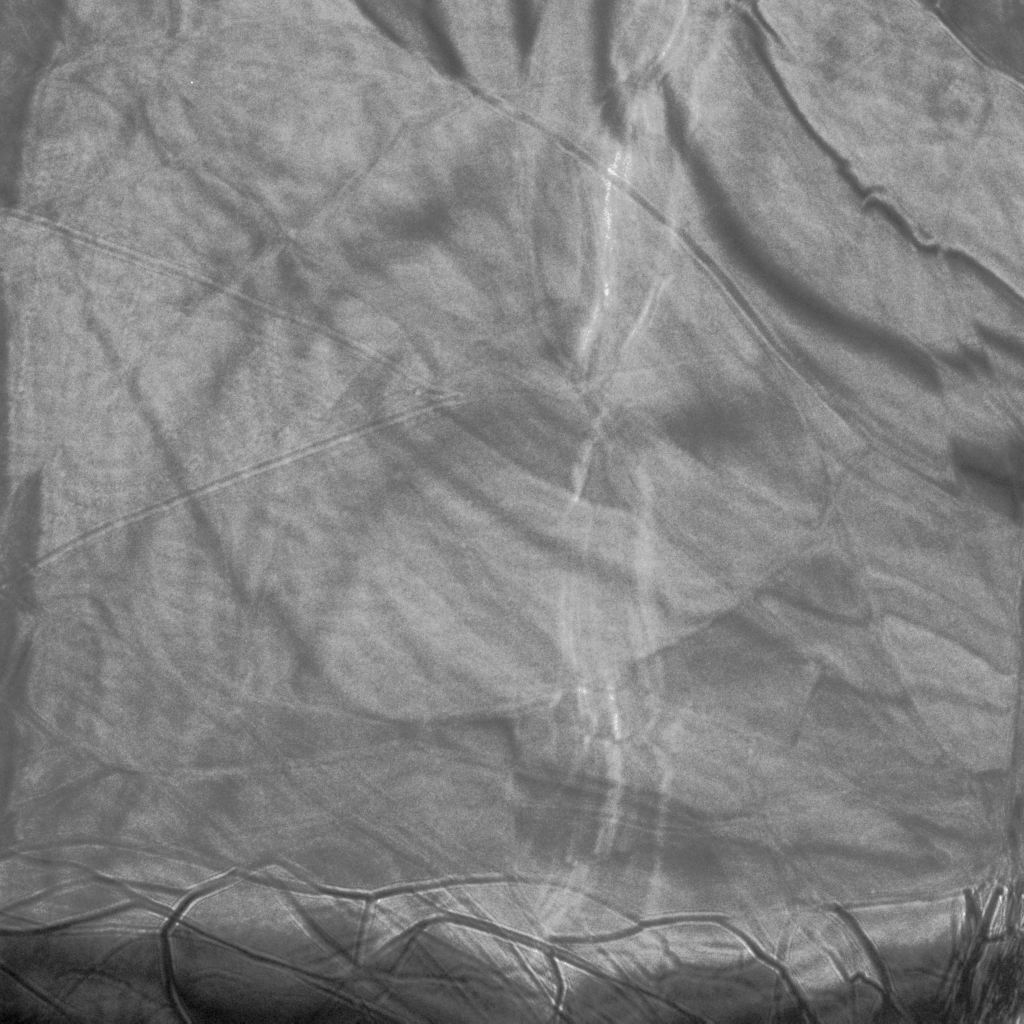}
		\caption{\Bc = 210 Oe}
	\end{subfigure}
	\begin{subfigure}[t]{0.3\textwidth}
		\centering
		\includegraphics[width=\textwidth]{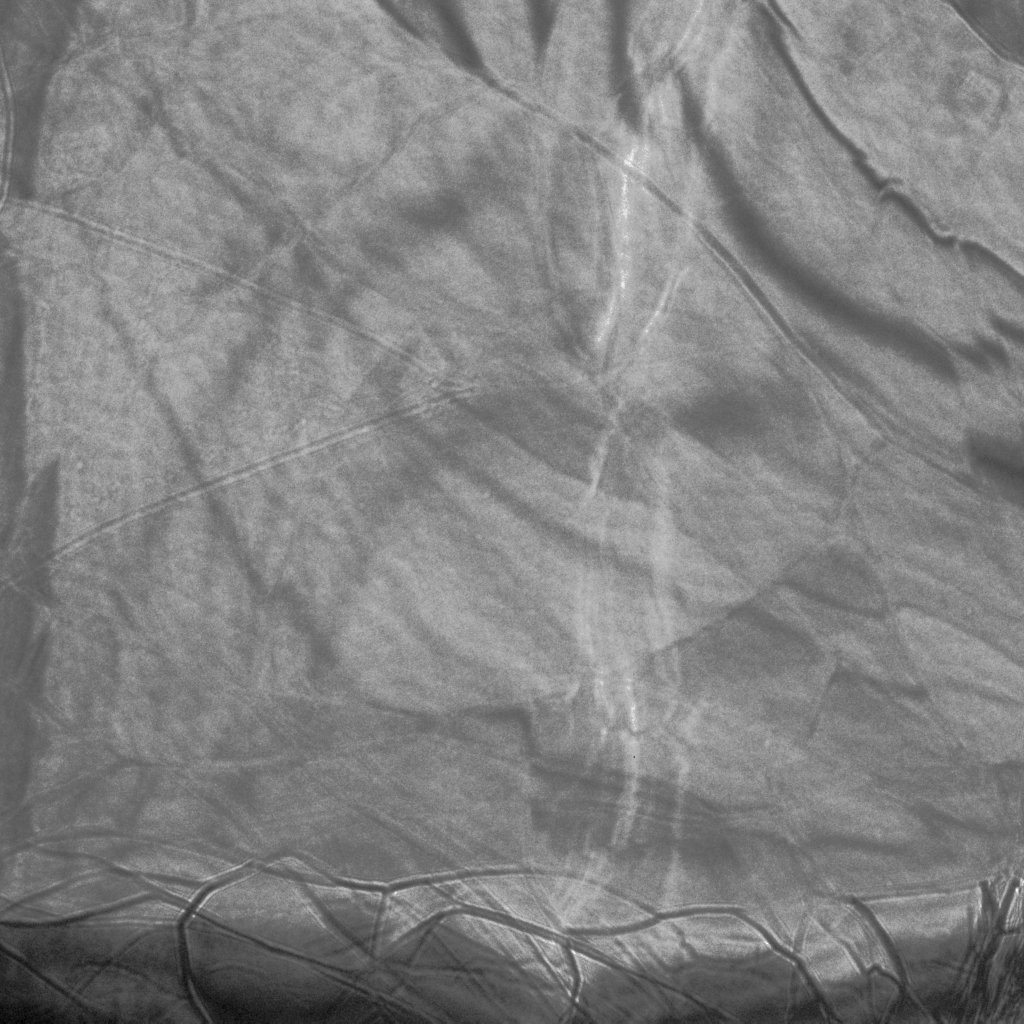}
		\caption{\Bc = 262 Oe}
	\end{subfigure}
	\begin{subfigure}[t]{0.3\textwidth}
		\centering
		\includegraphics[width=\textwidth]{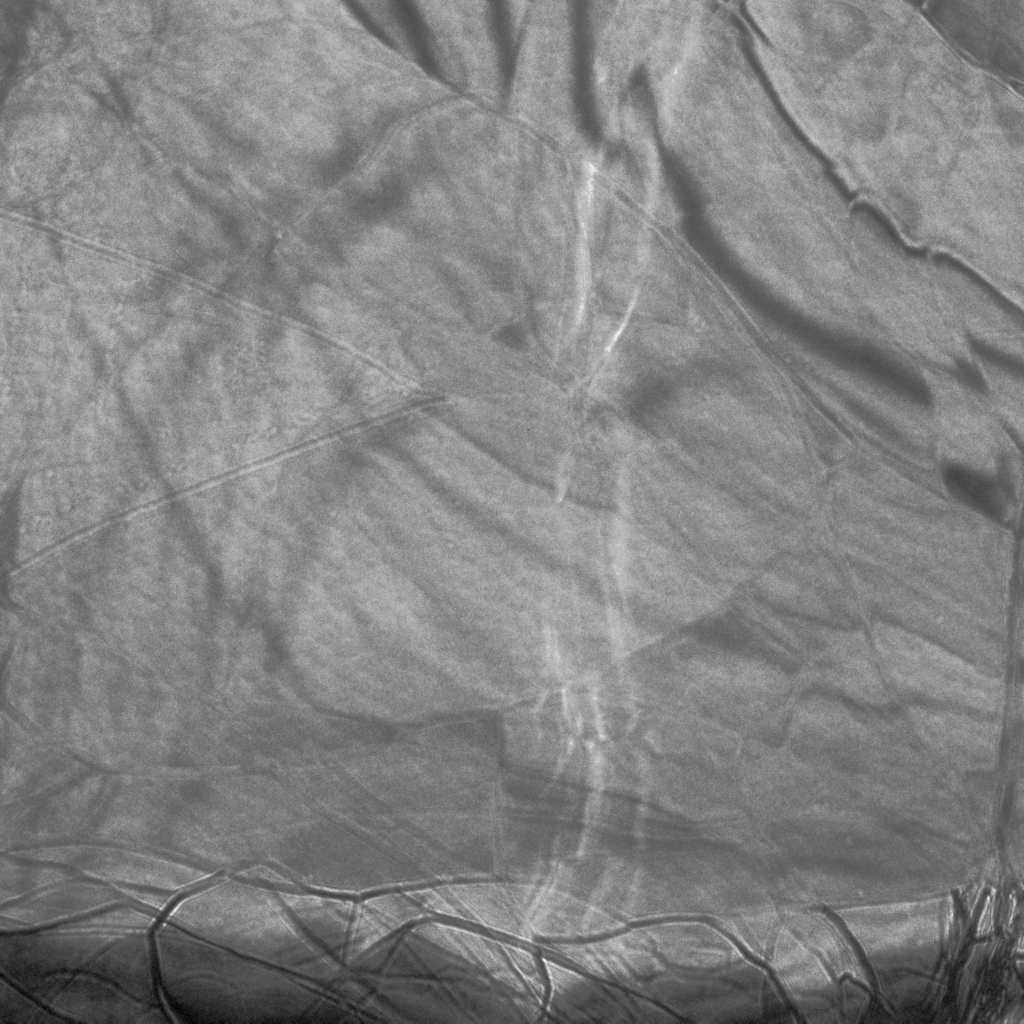}
		\caption{\Bc = 350 Oe}
	\end{subfigure}
	\begin{subfigure}[t]{0.3\textwidth}
		\centering
		\includegraphics[width=\textwidth]{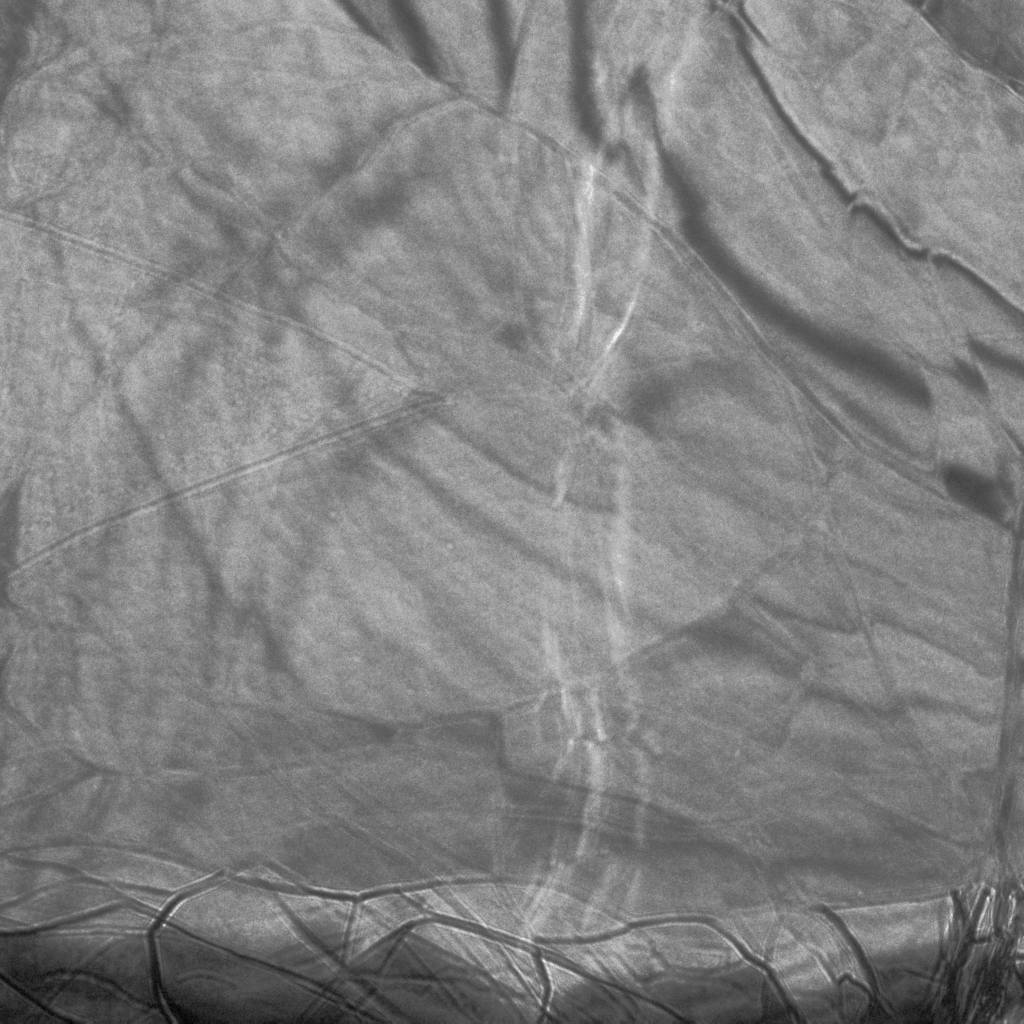}
		\caption{\Bc = 700 Oe}
	\end{subfigure}
	\begin{subfigure}[t]{0.3\textwidth}
		\centering
		\includegraphics[width=\textwidth]{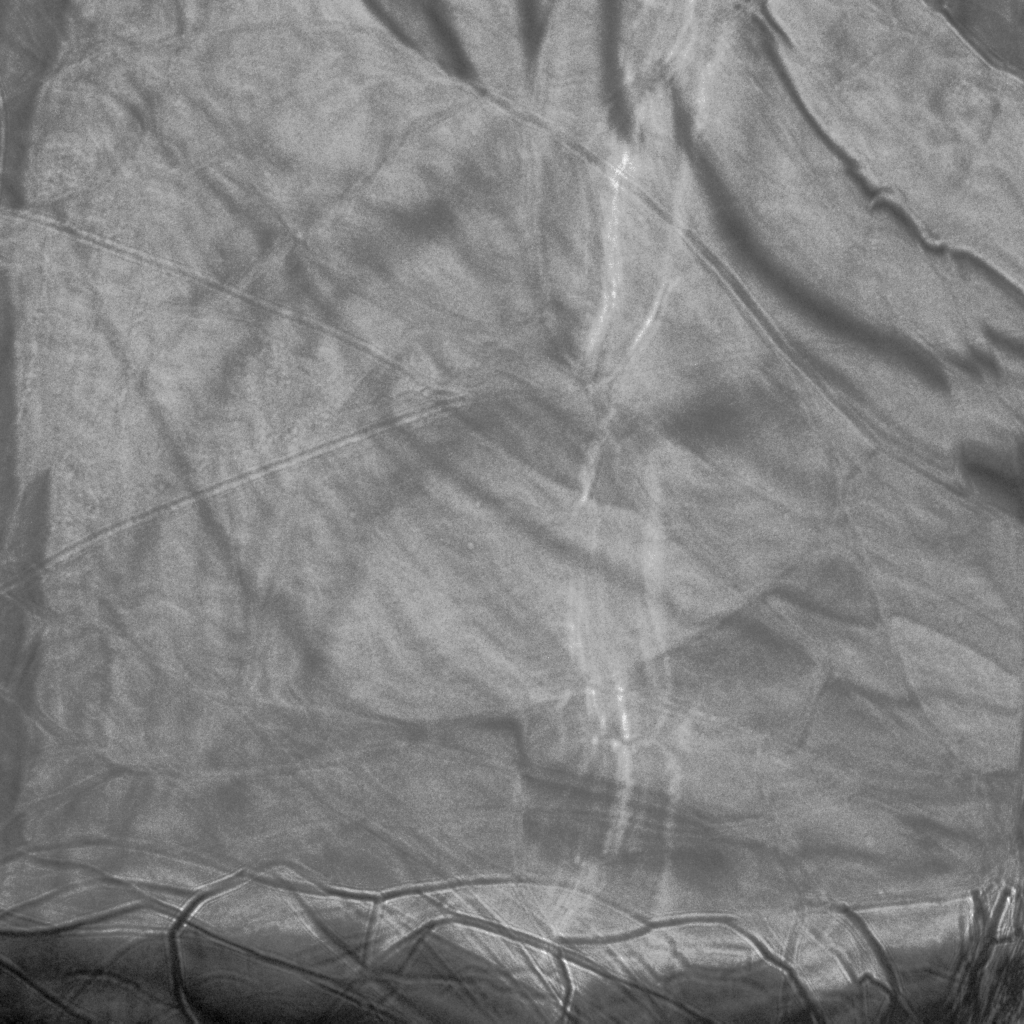}
		\caption{\Bc = 1050 Oe}
	\end{subfigure}
	\caption{T = 303K, image field of view is 5.97$\mu$m}
	\label{fig: PDinfo303K}
\end{figure}

\begin{figure}
	\centering
	\begin{subfigure}[t]{0.3\textwidth}
		\centering
		\includegraphics[width=\textwidth]{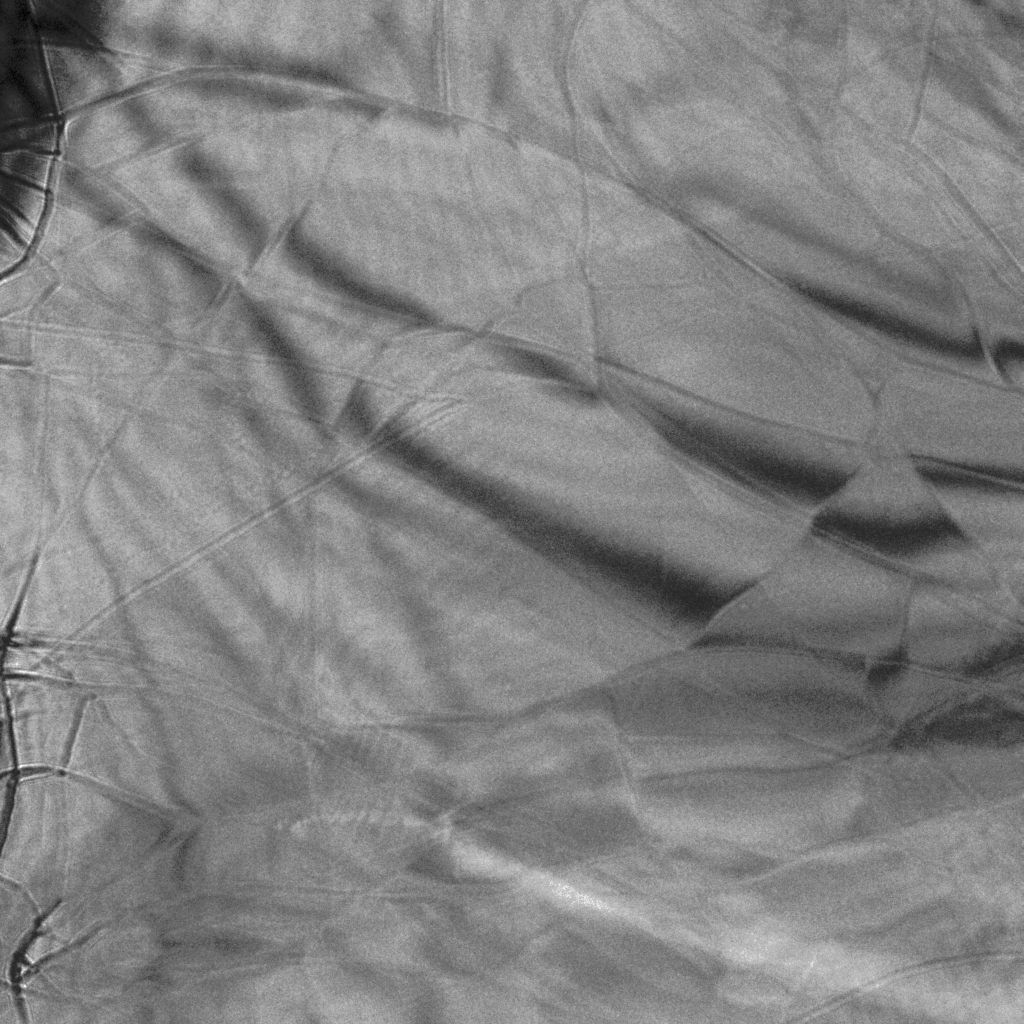}
		\caption{\Bc = 35 Oe}
	\end{subfigure}
	\begin{subfigure}[t]{0.3\textwidth}
		\centering
		\includegraphics[width=\textwidth]{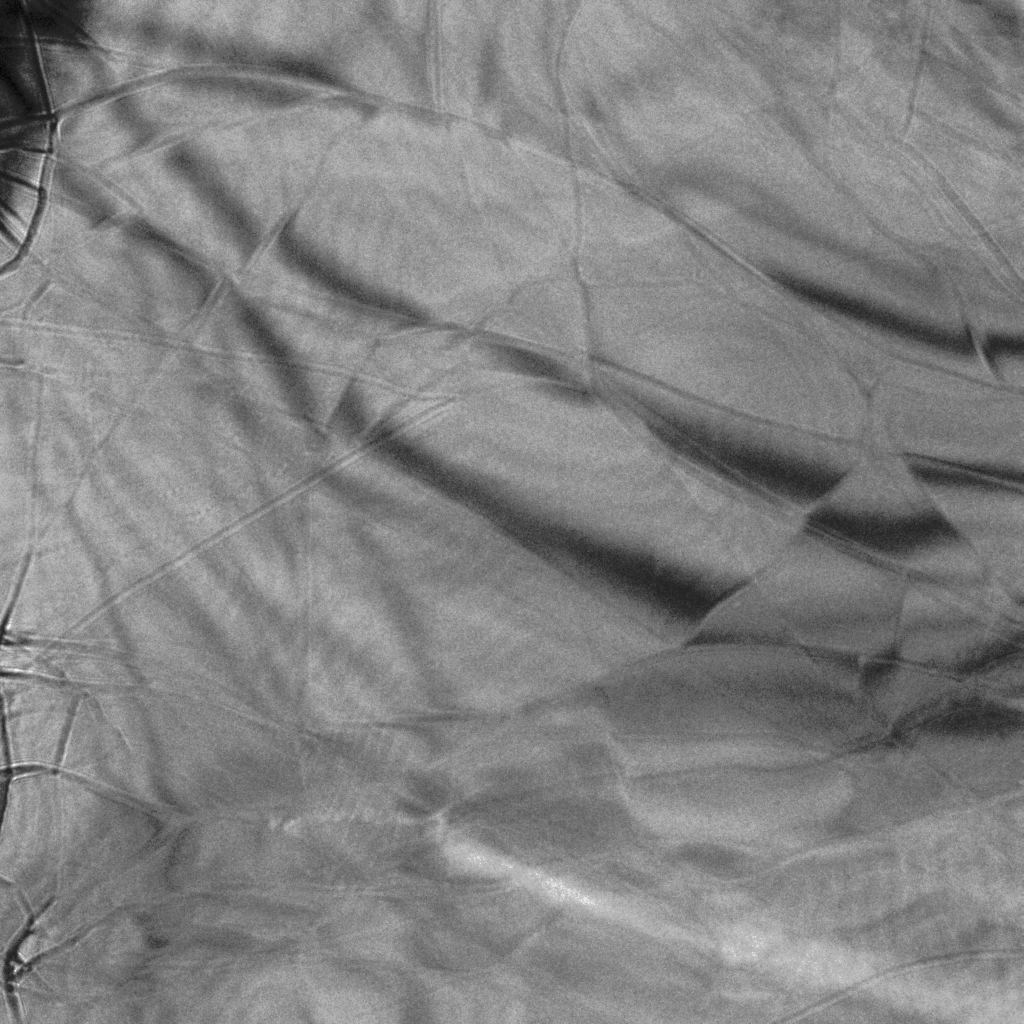}
		\caption{\Bc = 70 Oe}
	\end{subfigure}
	\begin{subfigure}[t]{0.3\textwidth}
		\centering
		\includegraphics[width=\textwidth]{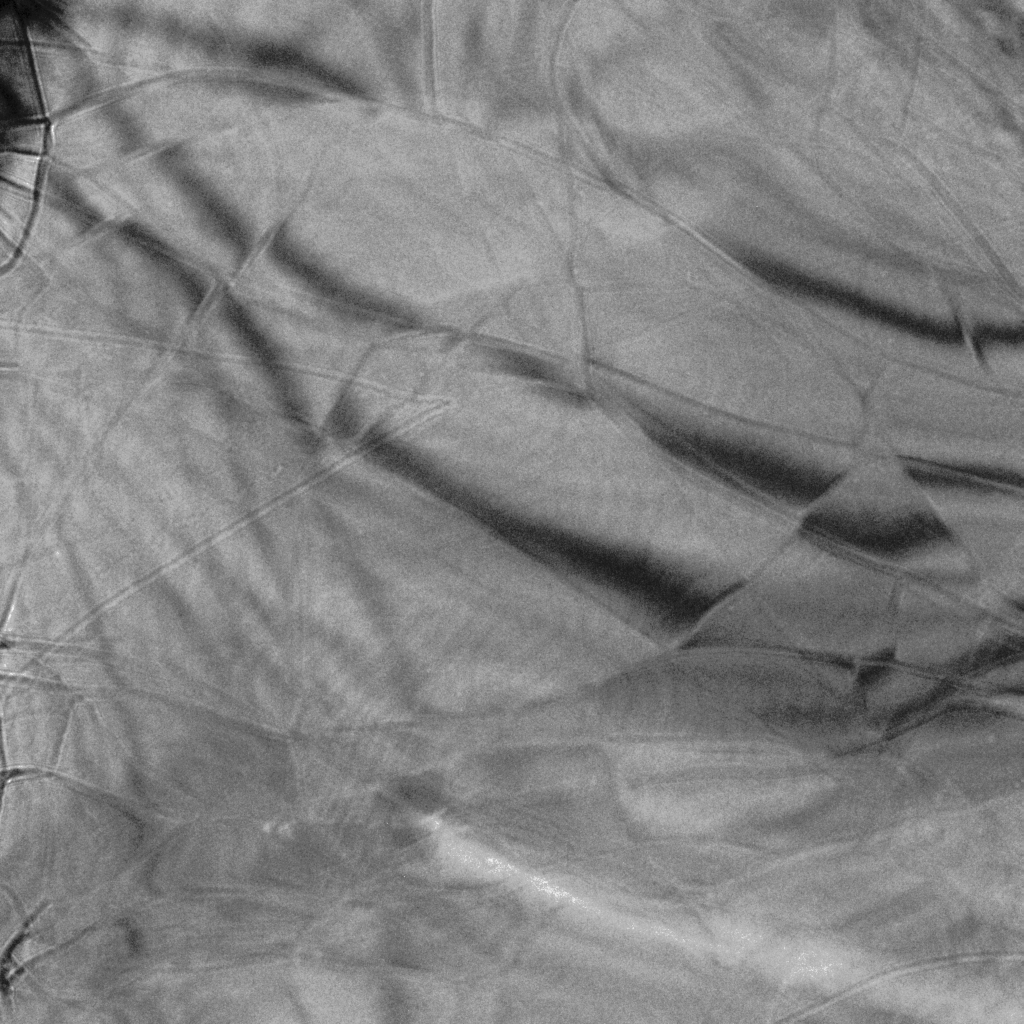}
		\caption{\Bc = 175 Oe}
	\end{subfigure}
	\begin{subfigure}[t]{0.3\textwidth}
		\centering
		\includegraphics[width=\textwidth]{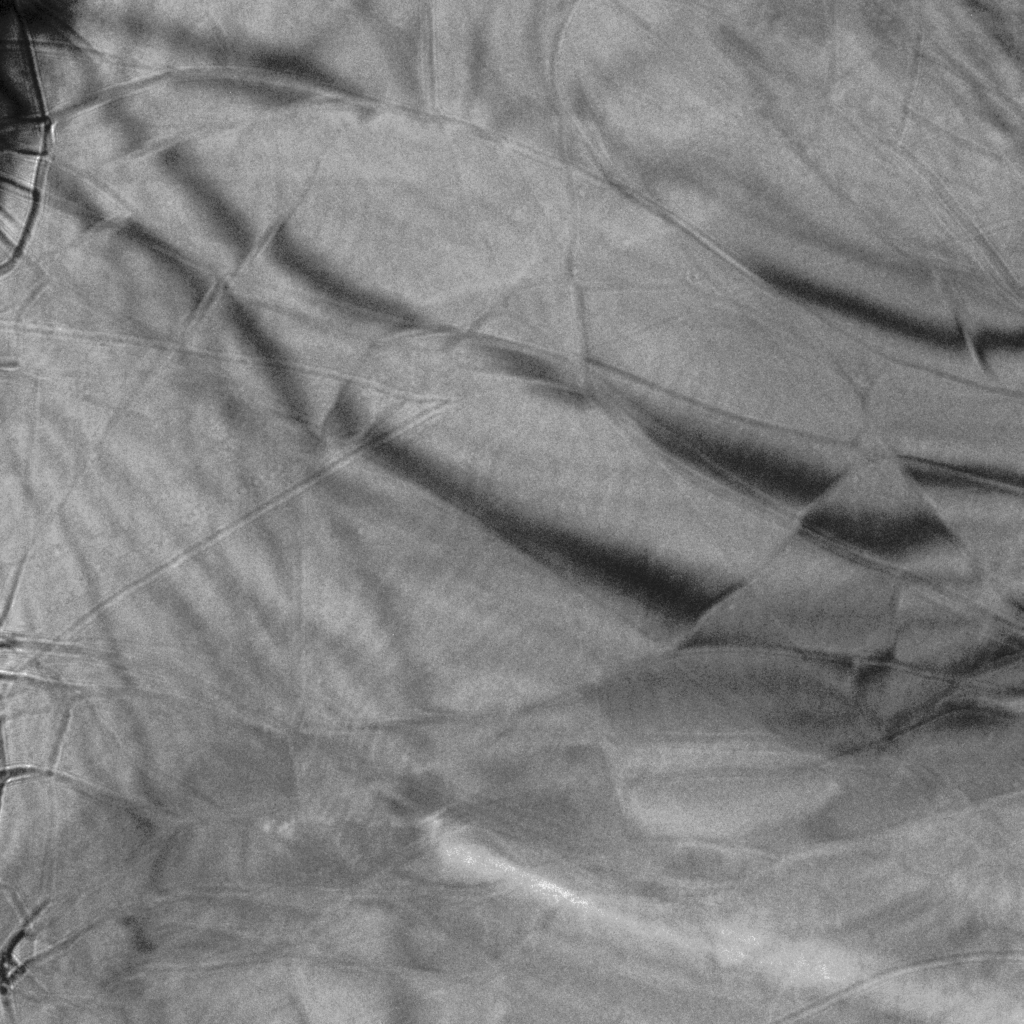}
		\caption{\Bc = 210 Oe}
	\end{subfigure}
	\begin{subfigure}[t]{0.3\textwidth}
		\centering
		\includegraphics[width=\textwidth]{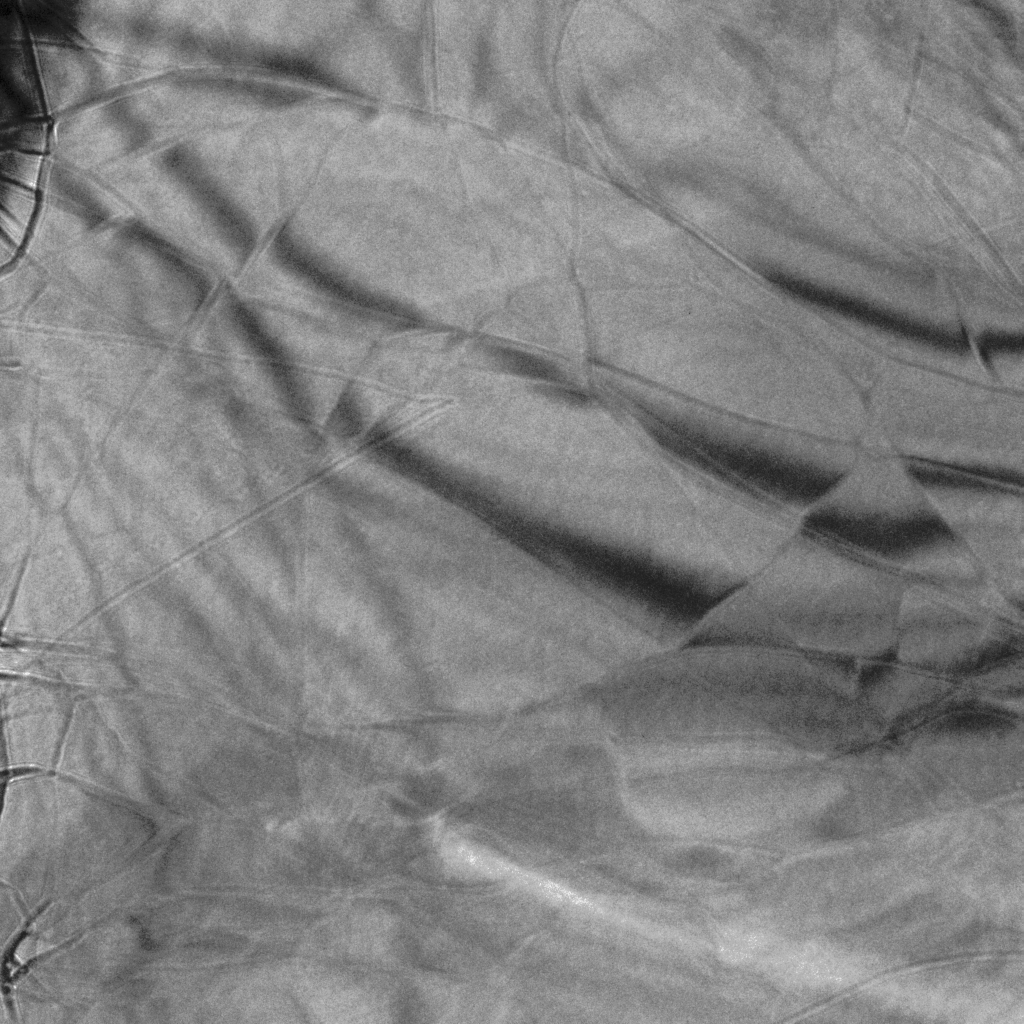}
		\caption{\Bc = 227 Oe}
	\end{subfigure}
	\begin{subfigure}[t]{0.3\textwidth}
		\centering
		\includegraphics[width=\textwidth]{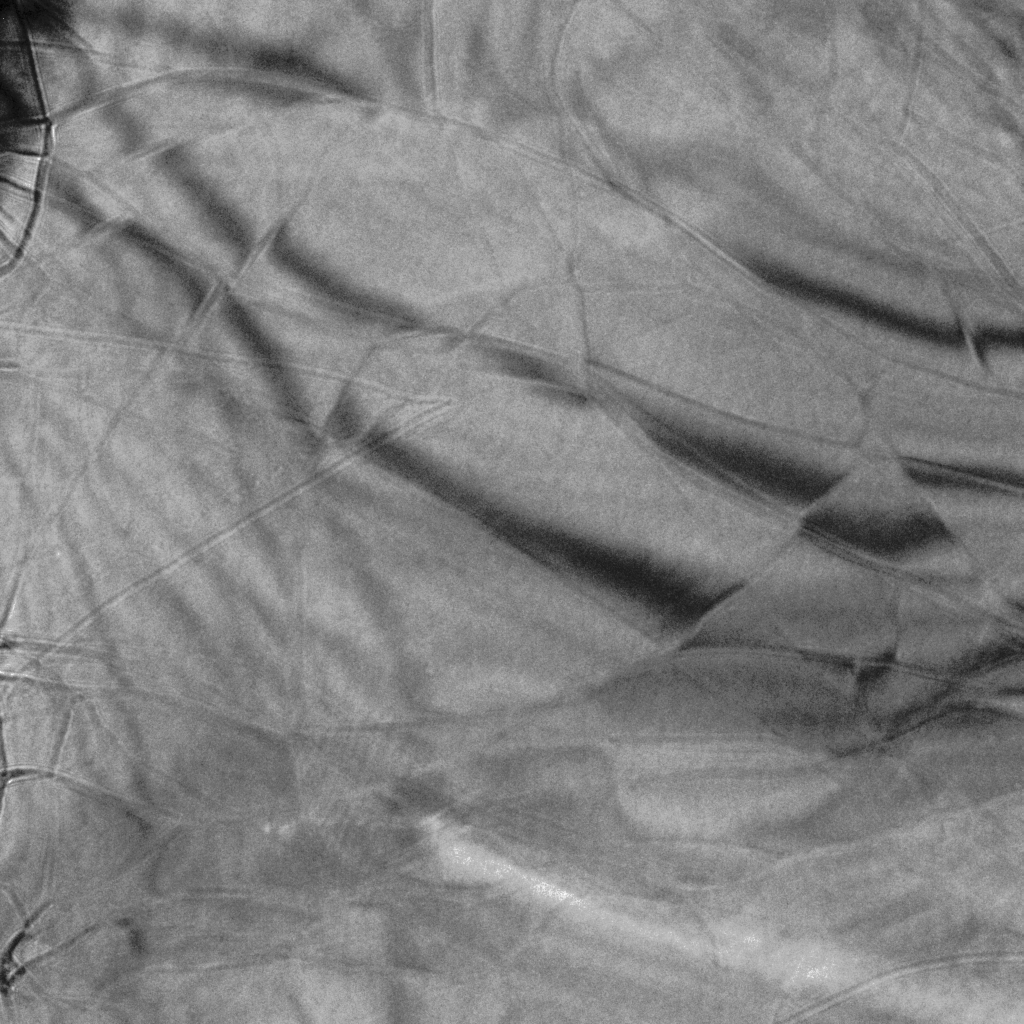}
		\caption{\Bc = 262 Oe}
	\end{subfigure}
	\begin{subfigure}[t]{0.3\textwidth}
		\centering
		\includegraphics[width=\textwidth]{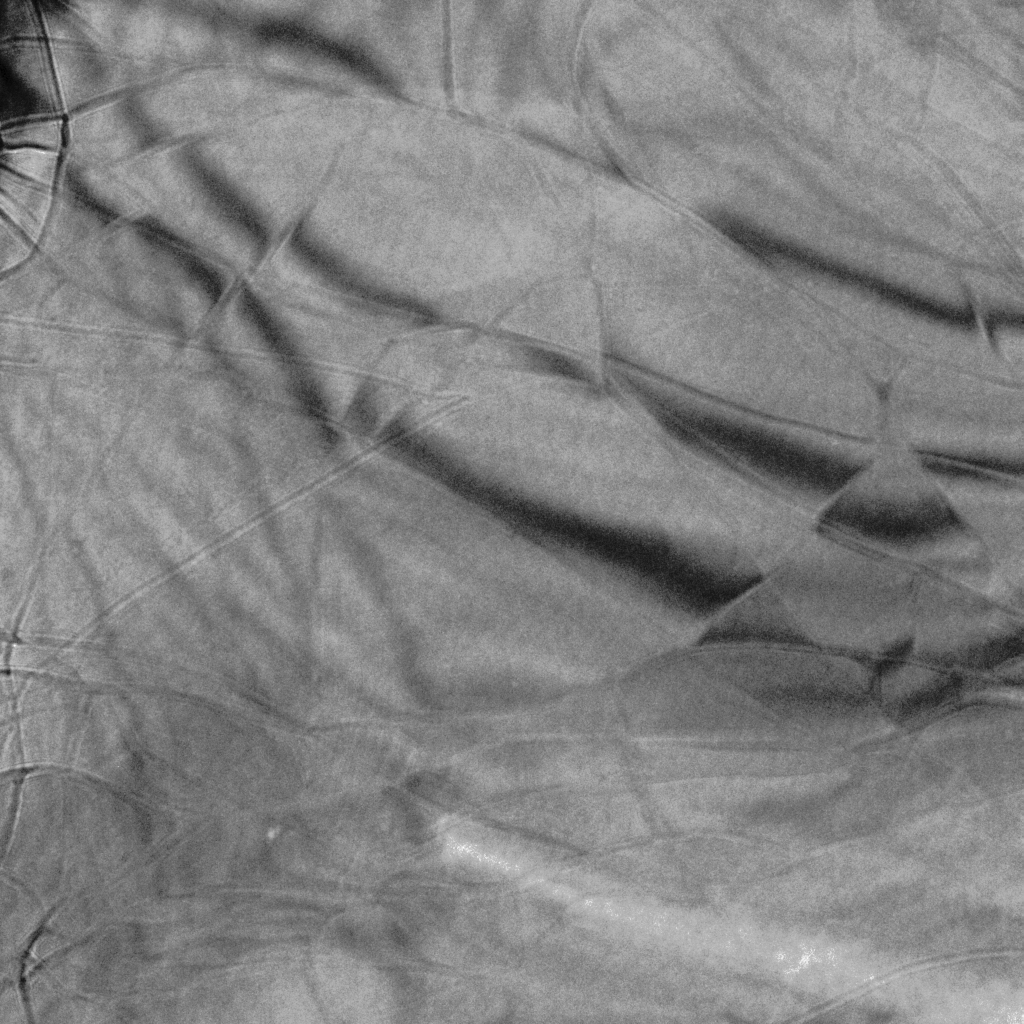}
		\caption{\Bc = 350 Oe}
	\end{subfigure}
	\begin{subfigure}[t]{0.3\textwidth}
		\centering
		\includegraphics[width=\textwidth]{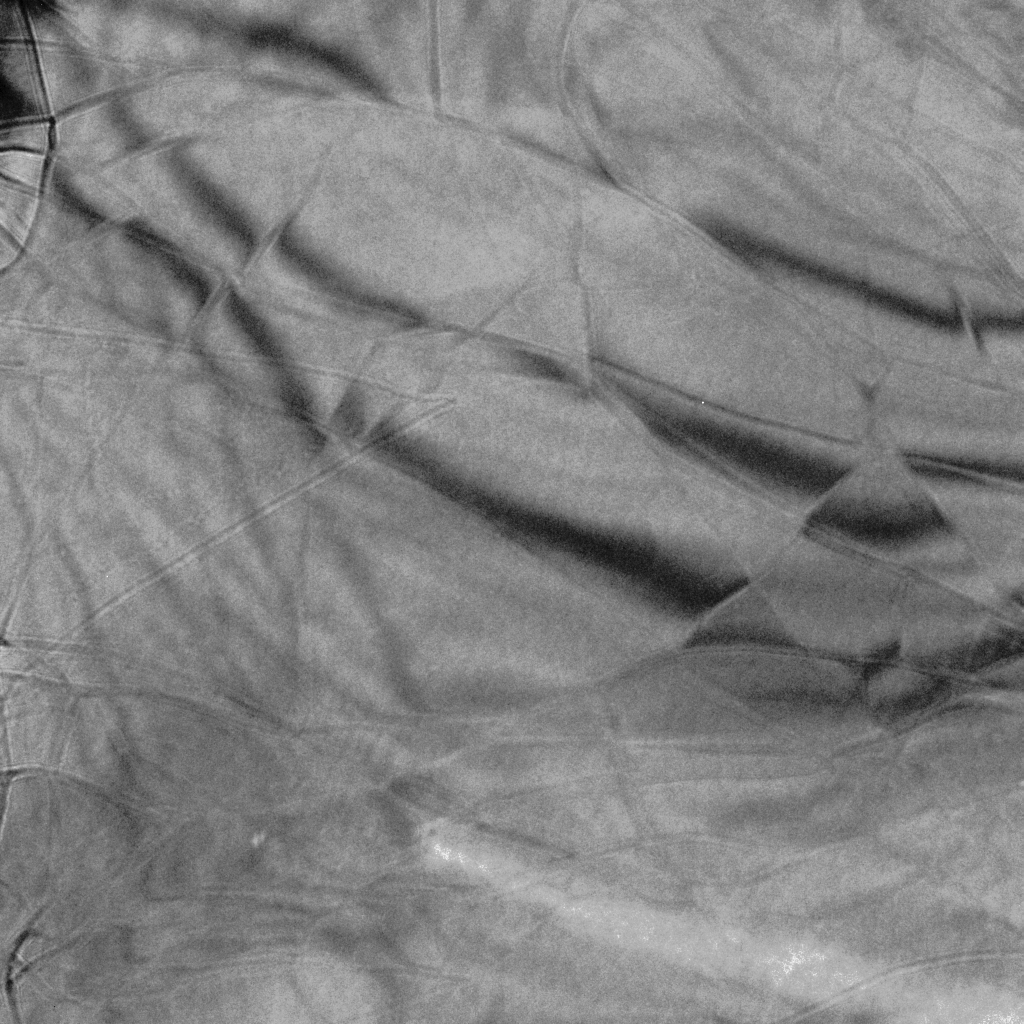}
		\caption{\Bc = 700 Oe}
	\end{subfigure}
	\begin{subfigure}[t]{0.3\textwidth}
		\centering
		\includegraphics[width=\textwidth]{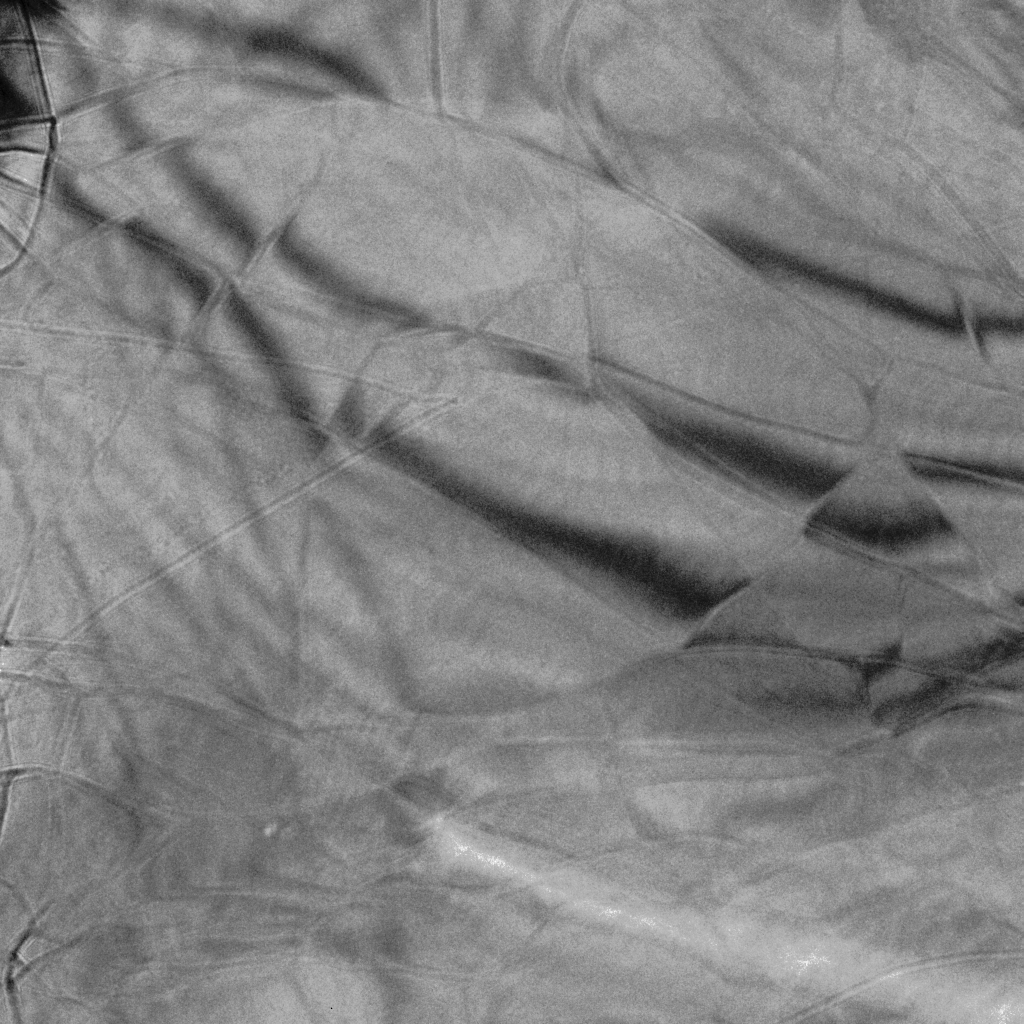}
		\caption{\Bc = 1050 Oe}
	\end{subfigure}
	\caption{T = 313K, image field of view is 5.97$\mu$m}
	\label{fig: PDinfo313K}
\end{figure}

\begin{figure}
	\centering
	\begin{subfigure}[t]{0.3\textwidth}
		\centering
		\includegraphics[width=\textwidth]{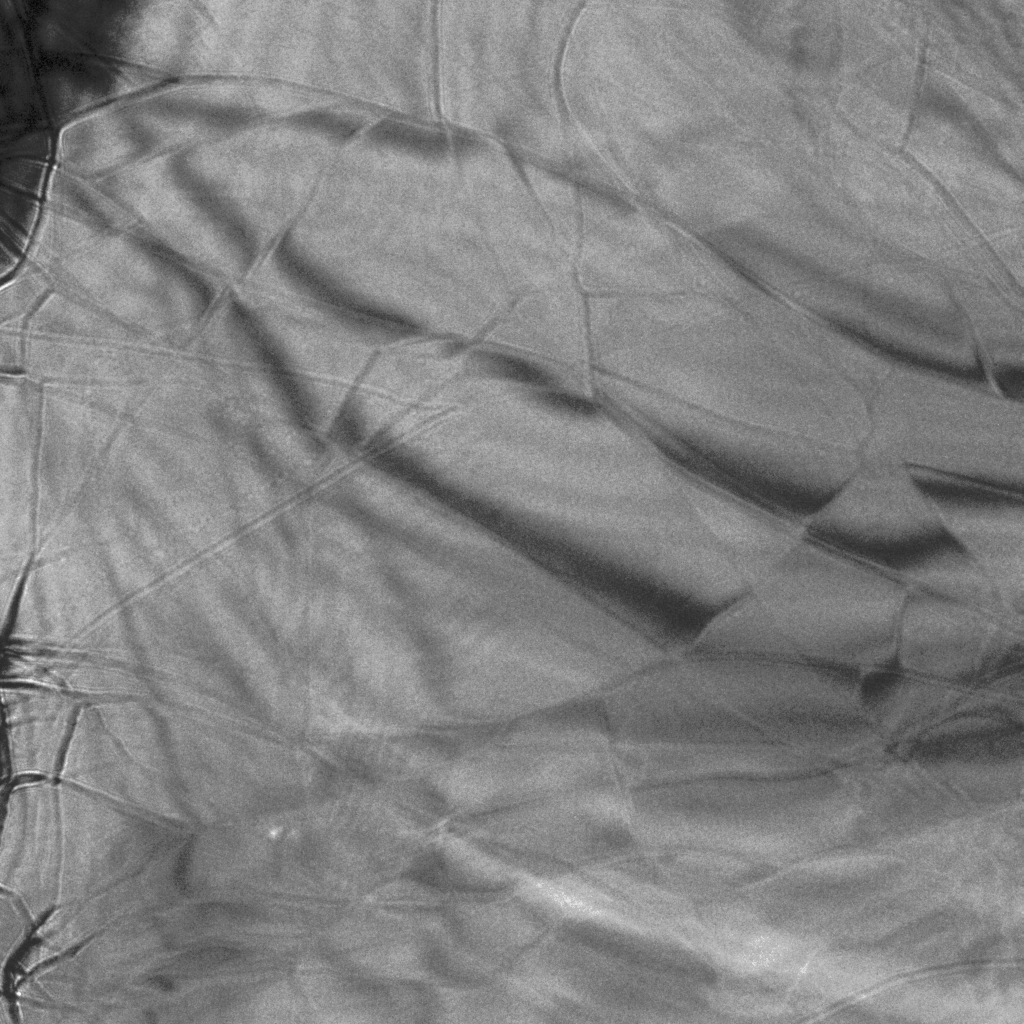}
		\caption{\Bc = 0 Oe}
	\end{subfigure}
	\begin{subfigure}[t]{0.3\textwidth}
		\centering
		\includegraphics[width=\textwidth]{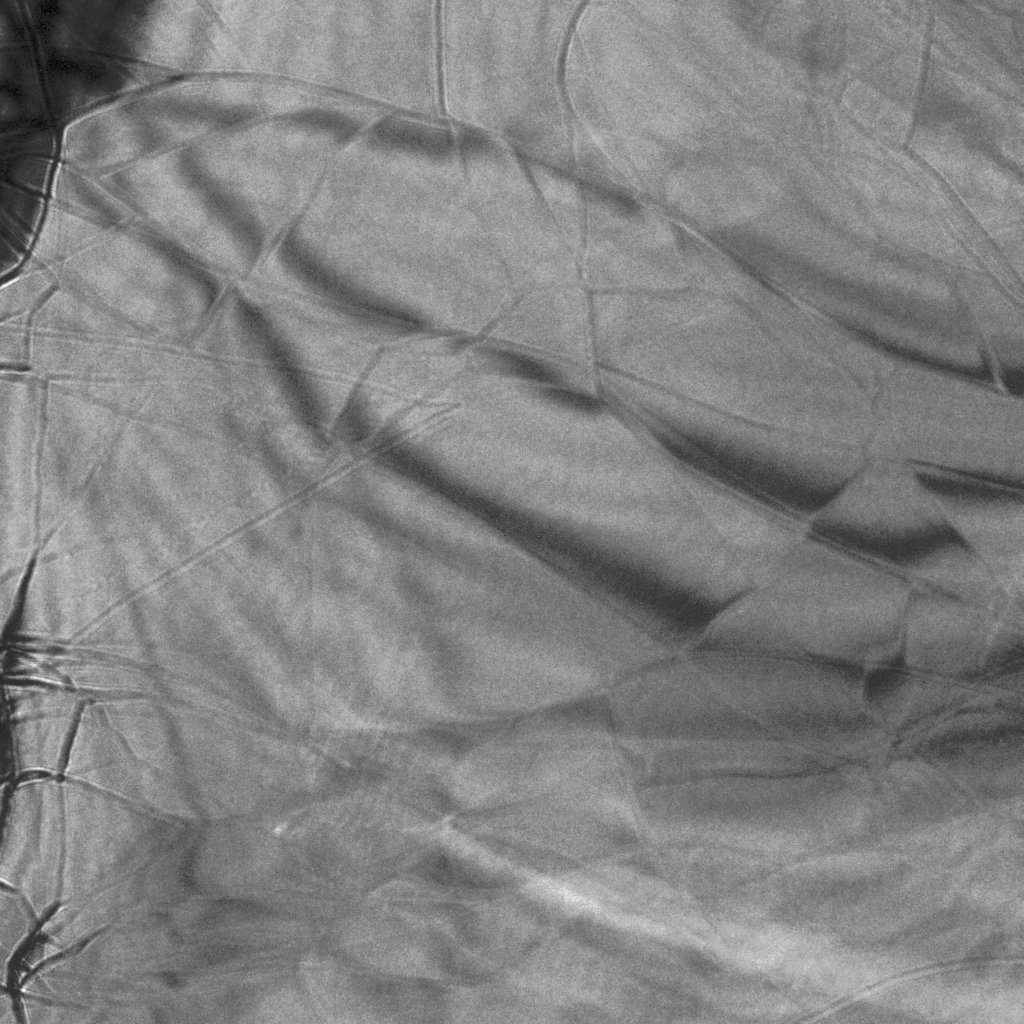}
		\caption{\Bc = 350 Oe}
	\end{subfigure}
	\begin{subfigure}[t]{0.3\textwidth}
		\centering
		\includegraphics[width=\textwidth]{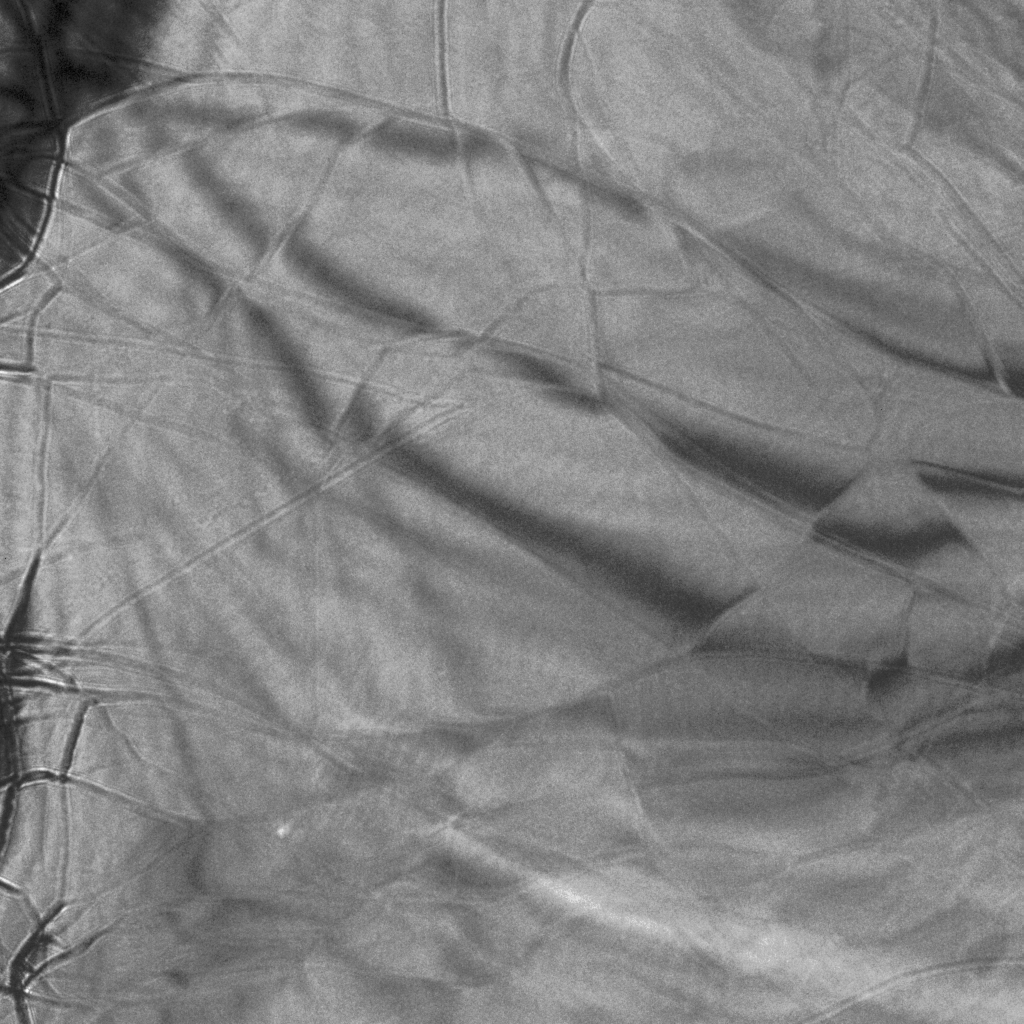}
		\caption{\Bc = 700 Oe}
	\end{subfigure}
	\begin{subfigure}[t]{0.3\textwidth}
		\centering
		\includegraphics[width=\textwidth]{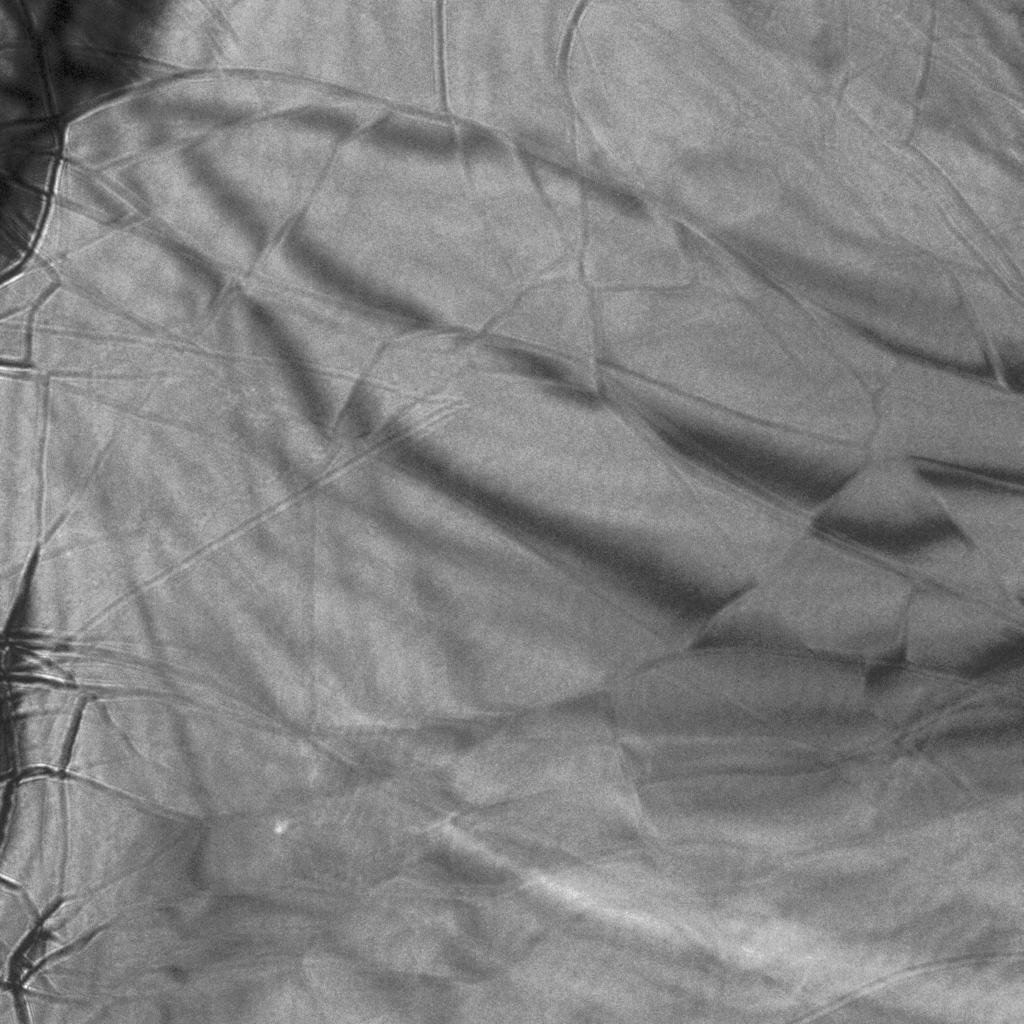}
		\caption{\Bc = 1050 Oe}
	\end{subfigure}
	\caption{T = 323K, image field of view is 5.97$\mu$m}
	\label{fig: PDinfo323K}
\end{figure}

Each data point on the phase diagram in Figure 3 comes from a LTEM image collected after a specific temperature and magnetic field treatment, as described previously. Figures \ref{fig: PDinfo200K}, \ref{fig: PDinfo250K}, \ref{fig: PDinfo273K}, \ref{fig: PDinfo293K}, \ref{fig: PDinfo303K}, \ref{fig: PDinfo313K}, \ref{fig: PDinfo323K} show the LTEM images used to classify the magnetic texture present after each treatment. Each image has a field of view of 5.97$\mu$m.

The data points at 1050 Oe cooling field and 203K, 253K, and 273K were inferred by applying a 1050 Oe magnetic field (higher than the saturation magnetic field) then returning the sample to zero magnetic field. This is equivalent to field cooling in 1050 Oe (See Figure 3). 


\clearpage

\bibliography{bibliography}